\documentclass[12pt,times]{nmepreprint} 

\usepackage[colorlinks, linktocpage=true, pdfpagemode=No,
linkcolor=blue, citecolor=blue, pagecolor=blue]{hyperref}

\usepackage{listings}
\usepackage{multirow}
\usepackage{csml}
\usepackage{float}
\usepackage{subfig}
\usepackage{booktabs}
\usepackage[table]{xcolor}
\usepackage{mathtools}
\usepackage{scalerel}
\usepackage{setspace}

\usepackage{algorithm}
\usepackage{algpseudocode}
\algtext*{EndWhile}
\algtext*{EndFor}
\algtext*{EndIf}

\graphicspath{{./figs/}}

\newcommand{\RVE}{\operatorname{\mathit{R\kern-.17em V\kern-.17em E}}}

\begin{document}

\runningheads{T. A. Archbold, I. Kazlauskaite and F. Cirak} {Multi-view Bayesian optimisation in reduced space}


\title{Multi-view Bayesian optimisation in an input-output reduced space for engineering design}

\author{Thomas A. Archbold\corrauth, Ieva Kazlauskaite and Fehmi Cirak}

\address
{Department of Engineering, University of Cambridge, Trumpington Street, Cambridge CB2 1PZ, U.K.
}

\corraddr{taa42@cantab.ac.uk}

\begin{abstract}
Bayesian optimisation is an adaptive sampling strategy for constructing a Gaussian process surrogate to efficiently search for the global minimum of a black-box computational model.
Gaussian processes have limited applicability in engineering design problems, which usually have many design variables but typically a low intrinsic dimensionality. Their scalability can be significantly improved by identifying a low-dimensional space of latent variables that serve as inputs to the Gaussian process.  In this paper, we introduce a multi-view learning strategy that considers both the input design variables and output data representing the objective or constraint functions, to identify a low-dimensional latent subspace. Adopting a fully probabilistic viewpoint, we use probabilistic partial least squares (PPLS) to learn an orthogonal mapping from the design variables to the latent variables using training data consisting of inputs and outputs of the black-box computational model. The latent variables and posterior probability densities of the PPLS and Gaussian process models are determined sequentially and iteratively, with retraining occurring at each adaptive sampling iteration. We compare the proposed probabilistic partial least squares Bayesian optimisation (PPLS-BO) strategy with its deterministic counterpart, partial least squares Bayesian optimisation (PLS-BO), and classical Bayesian optimisation, demonstrating significant improvements in convergence to the global minimum. 
\end{abstract}


\maketitle

%

%
\section{Introduction}\label{Introduction}

Conventional approaches to design optimisation require the gradients of objective and constraint functions such as the compliance or maximum stress, with respect to design variables often defined in terms of computer aided design (CAD) model parameters~\cite{Cirak:2002aa,hughes_isogeometric_2005,bandara_2016_shape,lieu2017multiresolution,yin_2020_topologically,xiao2022infill}. In practice such gradient information may not be available and could quickly become costly to evaluate, especially for multiphysics problems with complex geometry. Consequently, it is expedient to interpret the computational model as a black-box function where output quantities of interest (QOIs), such as objective and constraint function values, are observed for a finite set of prescribed training inputs (design variables). Output QOIs for arbitrary inputs are inferred using a quick-to-evaluate surrogate that emulates the black-box function. Clearly, a surrogate cannot precisely reproduce a black-box function and uncertainty about the inferred QOI is unavoidable.   Statistical surrogates such as Gaussian processes (GPs)~\cite{williams2006gaussian,forrester2008engineering,diazdelao2011gaussian} yield a probability density to quantify the uncertainty in inferring unobserved QOIs. Although GPs have been used effectively for engineering design~\cite{sakata2003structural,xiong2007non,bessa2018design}, they do not scale to problems involving a large number of input variables, i.e. they are susceptible to the "curse of dimensionality"~\cite{bengio_curse_2005}. GPs can be applied to design optimisation using adaptive sampling methods such as Bayesian optimisation (BO)~\cite{jones_efficient_1998,fuhg2021state}. 

In this paper, we emulate each QOI defined via the black-box computational model, using a GP surrogate $f(\vec{s})$. The Gaussian prior probability density characterised by its mean and covariance, is updated using a training data set $\mathcal{D}$ and Bayes' rule to yield the posterior probability density. The training data set~$\mathcal{D} = \{(\vec{s}_i,y_i)\}_{i=1}^n$ collects the $n$ pairs of design variables $\vec{s}_i \in \mathbb{R}^{d_s}$ and the QOI $y_i \in \mathbb{R}$  from the computational model. The likelihood function follows from the posited statistical model $y = f(\vec{s}) + \epsilon_y$, with an additive noise (error) term $\epsilon_y$. We use BO to perform global optimisation to find optimum design variables that minimise the objective function and satisfy the constraints~\cite{horst2013handbook}. BO is an adaptive sampling methodology that iteratively updates the GP using samples proposed by an acquisition function~\cite{frazier2018tutorial}. The acquisition function balances the exploration of the design space with the exploitation of regions where the GP posterior predicts the global minimum~\cite{wilson2018maximizing}. The training data set $\mathcal{D}$ expands with each proposed sample, and the GP is updated to fit the new data. Like standard GPs, BO is prone to the "curse of dimensionality" and is typically limited to engineering design problems with approximately~$10$ or fewer design parameters, see e.g.~\cite{mathern_multi-objective_2021,morita2022applying} and~\cite{sheikh2022optimization}. 

For computational tractability when using GP surrogates, the dimension~$d_s$ of the design variable vector~$\vec{s}$ must be restricted. To achieve this, we assume that the GP $f(\vec{z})$ depends on a low-dimensional latent variable vector~$\vec z$  rather than $\vec s$. The latent variables are obtained from a linear mapping $\vec{z} = \vec{W}^\trans\vec{s}$, with a non-square (tall) orthogonal matrix~$\vec W$,  i.e.~$\vec{W}^\trans\vec{W} = \vec{I}$. The orthonormal rows and columns of the matrix~$\vec W$ represent the low-dimensional latent subspace corresponding to~$\vec z$.  Engineering design problems usually consist of multiple QOIs, such as the compliance objective function and stress constraints; both these must be considered in constructing the orthogonal matrix~$\vec W$. We adopt a multi-view learning (MVL) strategy to discover the low-dimensional orthonormal basis~\cite{zhao_multi-view_2017}. Although the suite of MVL methods is extensive~\cite{xu2013survey}, sensible models are not deterministic but have epistemic uncertainties when predicting unobserved data (as is the case when computing unobserved latent variables)~\cite{ghahramani2015probabilistic}.  Epistemic uncertainties decrease with more training data and converge to zero in the limit of infinite data. 
 
In the proposed approach, we adopt a fully probabilistic viewpoint by combining BO with probabilistic partial least squares (PPLS). PPLS~\cite{el2018probabilistic} postulates a probabilistic model given by~\mbox{$\vec{s} = \vec{W}\vec{z} + \hat{\vec{\epsilon}}_s$} and~\mbox{$\vec{y} = \vec{Q}\vec{z} + \hat{\vec{\epsilon}}_y$}, such that  design variable and QOI vector pairs~\mbox{$(\vec{s}, \vec{y})$} are generated from an unobserved latent variable vector~$\vec{z}$. The error incurred by reconstructing~$\vec s$ and~$\vec y$ from~$\vec z$ is assumed to have Gaussian probability densities, i.e.~$\hat{\vec{\epsilon}}_s \sim \mathcal{N}(\vec{0},\hat{\vec{\Sigma}}_s)$ and~$\hat{\vec{\epsilon}}_y \sim \mathcal{N}(\vec{0},\hat{\vec{\Sigma}}_y)$. The entries of the orthogonal matrices $\vec{W}$ and $\vec{Q}$, and covariance matrices $\hat{\vec{\Sigma}}_s$ and $\hat{\vec{\Sigma}}_y$ are treated as PPLS model hyperparameters. The hyperparameters and the approximate posterior probability density of the latent variable vector~$\vec z$ are determined using variational Bayes~\cite{jordan1999introduction}. We propose probabilistic partial least squares Bayesian optimisation (PPLS-BO) that combines PPLS with adaptive sampling using BO. The probabilistic model ensures that each latent variable proposed by the BO acquisition function yields a density of corresponding design variables. By sampling from this density, we show that exploration is enhanced, leading to improved convergence towards the global minimum and increased robustness to misspecification in the dimension of the low-dimensional latent space.  

BO approaches for high-dimensional problems have received substantial attention in the machine learning and engineering communities. Trust region Bayesian optimisation (TuRBO) fits a GP within a subdomain and solves a local optimisation problem, expanding the subdomain to balance exploration with exploitation~\cite{eriksson2019scalable}. However, TuRBO does not restrict exploration to the low intrinsic dimensionality present in many engineering problems. Where the problem has a low intrinsic dimensionality, adaptive sampling efficiency can be improved by learning the low-dimensional subspace. If the intrinsic dimensionality varies across the domain, restricting exploration to a single subspace can prevent the optimiser from finding the optimum; in such cases, local optimisation strategies such as TuRBO may be more effective. Many approaches for learning a low-dimensional subspace are based on deep-learning methods~\cite{tripp2020sample,notin2021improving} that require large quantities of data that are computationally expensive to obtain for complex engineering designs. Alternative methods such as principal component analysis (PCA)~\cite{berkooz_proper_1993,dulong_model_2007} or random embedding~\cite{wang_bayesian_2013}, which discover a low-dimensional subspace based solely on the input design variables alone, are not appropriate in this context since they do not take into account the optimisation objectives and constraints. Alternatively, gradient information about the QOI can be used to discover a low-dimensional "active" subspace~\cite{constantine_active_2014,li_surrogate-based_2019,lam_multifidelity_2020,romor2021multi}. Where gradient information is unavailable, the input design variable covariance can be weighted using output data~\cite{raponi_high_2020}, or interrogated when combined with GPs using a penalised log-likelihood formulation~\cite{gaudrie_modeling_2020}. The subspace basis can be optimised jointly with the GP hyperparameters using two-step algorithms based on coordinate descent~\cite{tripathy2016gaussian,tsilifis2021bayesian}. Partial least squares (PLS) has been combined with GPs~\cite{song_novel_2012,bouhlel_improving_2016,zhu2019energy,zuhal2021dimensionality} and extended to BO for adaptive sampling with multiple QOIs~\cite{amine2018efficient,bartoli2019adaptive}. History-matching techniques similarly use sequential Bayesian updates to refine reduced subspaces with adaptive sampling guided by implausibility screening~\cite{vernon2010galaxy,vernon2018bayesian,salter2019uncertainty}. We compare our approach PPLS-BO, to its deterministic counterpart PLS-BO, and to BO performed over a low-dimensional subspace learned using PCA, which we abbreviate in this paper as PCA-BO for brevity.

This paper is structured as follows. In Section~\ref{sec:review}, we review PPLS for dimensionality reduction and  BO, consisting of GPs and acquisition functions for adaptive sampling. In Section~\ref{sec:adaptive_reduced}, we present the proposed PPLS-BO algorithm for adaptive sampling in reduced dimension. We detail the formulations used to compute the posterior probability density of the GP in reduced dimension, and provide the pseudocode of the PPLS-BO algorithm. Three examples are given in Section~\ref{sec:examples}, demonstrating the improved convergence of PPLS-BO when compared to PLS-BO and classical BO. We demonstrate the algorithms' versatility through design optimisation of a complex manufacturing example. Finally, Section~\ref{sec:conclusion} concludes the paper and discusses promising directions for further research.
\section{Background}\label{sec:review}

In this section, we provide the relevant background on design optimisation, PPLS, and BO that are used in the proposed approach.

\subsection{Design optimisation}\label{subsec:design_opt}

In structural optimisation, the objective and constraint functions are determined by solving a FE model. The objective $J: \mathbb{R}^{d_s} \rightarrow \mathbb{R}$ and constraint $H^{(j)}: \mathbb{R}^{d_s} \rightarrow \mathbb{R}$ are functions of the design variables $\vec{s} \in \mathbb{R}^{d_s}$. A total of $d_y$ objective and constraint functions, consisting (without loss of generality) of a single objective and $d_y-1$ constraint functions, are considered. The design optimisation problem may be stated as
\begin{equation}\label{eq:ddo_1}
\begin{aligned}
\min_{s \in D_s} \quad & J(\vec{s}),\\
\textrm{s.t.} \quad & H^{(j)}(\vec{s}) \leq 0,  & \quad \quad  j&\in\{1,\,2,\,\cdots,\,d_y-1\},\\
  & D_s = \{\vec{s} \in \mathbb{R}^{d_s} \, \vert \, \bar{\vec{s}}^{(l)} \leq \vec{s} \leq \bar{\vec{s}}^{(u)}\},
\end{aligned}
\end{equation}
where $D_s \subset R^{d_s}$ is the design domain, bounded between a lower limit $\bar{\vec{s}}^{(l)}$ and upper limit $\bar{\vec{s}}^{(u)}$. We collect the objective and constraint function evaluations in an output vector~\mbox{$\vec{y} \in \mathbb{R}^{d_y}$ such that~\mbox{$\vec{y} = (J(\vec{s})\,\,H^{(1)}(\vec{s})\,\,H^{(2)}(\vec{s})\cdots H^{(d_y - 1)}(\vec{s}))^\trans$}}.

\subsection{Probabilistic partial least squares}\label{subsec:ppls}

\begin{figure}[t!]
\centering
{\includegraphics[width=80mm]{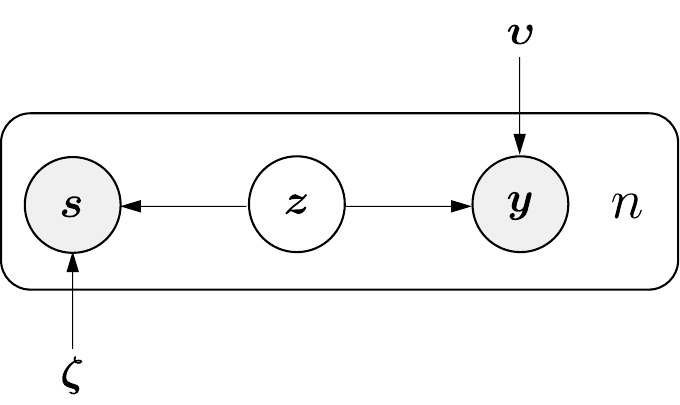}}
\caption{Graphical model of PPLS where $n$ is the size of the training data set $\mathcal{D}$, $\vec{s}$ are the design variables, $\vec{z}$ are the low-dimensional latent variables, $\vec{y}$ are the observations, and $\vec{\zeta}$ and $\vec{\upsilon}$ are two vectors containing the PPLS model hyperparameters.}
\label{fig:ppls_graphical_model}
\end{figure}

PPLS extends deterministic PLS (Appendix \hyperlink{Appendix A}{A}) by treating the low-dimensional latent space as uncertain~\cite{el2018probabilistic}. In Bayesian analysis, the uncertain latent variables are treated as random variables. PPLS is based on the following statistical observation (or, generative) model,
\begin{subequations}\label{eq:ppls_1}
\begin{align}
\vec{s} &= \vec{W}\vec{z} + \hat{\vec{\epsilon}}_s, \quad \quad \hat{\vec{\epsilon}}_s \sim \mathcal{N}\left(\vec{0},\hat{\vec{\Sigma}}_s\right),\label{eq:ppls_1a}\\
\vec{y} &= \vec{Q}\vec{z} + \hat{\vec{\epsilon}}_y, \quad \quad \hat{\vec{\epsilon}}_y \sim \mathcal{N}\left(\vec{0},\hat{\vec{\Sigma}}_y\right).\label{eq:ppls_1b}
\end{align}
\end{subequations}
The unobserved latent variable vector $\vec{z} \in \mathbb{R}^{d_z}$ is defined in a linear subspace given by the columns of the orthogonal matrices $\vec{W} \in \mathbb{R}^{d_s \times d_z}$ and $\vec{Q} \in \mathbb{R}^{d_y \times d_z}$, with $\vec{W}^\trans\vec{W} = \vec{Q}^\trans\vec{Q} = \vec{I}$. Reconstructing the design variables~$\vec s$ and outputs~$\vec y$ incur reconstruction errors modelled by zero-mean Gaussian random vectors $\hat{\vec{\epsilon}}_s$ and $\hat{\vec{\epsilon}}_y$ with diagonal covariance matrices $\hat{\vec{\Sigma}}_s \in \mathbb{R}^{d_s \times d_s}$ and $\hat{\vec{\Sigma}}_y \in \mathbb{R}^{d_y \times d_y}$ respectively. The joint probability density
\begin{equation}\label{eq:ppls_2}
\begin{aligned}
p_{\Phi}(\vec{y},\vec{s},\vec{z}) = p_\upsilon(\vec{y}\vert\vec{z}) p_\zeta(\vec{s}\vert\vec{z})p(\vec{z}),
\end{aligned}
\end{equation}
of all observed an unobserved variables is factorised using the posited conditional independence structure underlying \eqref{eq:ppls_1}, as also visualised in Figure \ref{fig:ppls_graphical_model}. According to~\eqref{eq:ppls_1a} and~\eqref{eq:ppls_1b} the prior probability densities for~$\vec s$ and~$\vec y$ are given by
\begin{subequations}\label{eq:ppls_4}
\begin{align}
p_\zeta(\vec{s}\vert\vec{z}) &= \mathcal{N}\left(\vec{W}\vec{z}, \hat{\vec{\Sigma}}_s\right),\label{eq:ppls_4a}\\
p_\upsilon(\vec{y}\vert\vec{z}) &= \mathcal{N}\left(\vec{Q}\vec{z}, \hat{\vec{\Sigma}}_y\right)\label{eq:ppls_4b}.
\end{align}
\end{subequations}
The trainable PPLS model hyperparameters are collected in the two sets $\vec{\zeta} = \{\vec{W},\hat{\vec{\Sigma}}_s\}$ and \mbox{$\vec{\upsilon} = \{\vec{Q},\hat{\vec{\Sigma}}_y\}$}. The prior probability density of the latent variable vector~$\vec z$ is assumed as 
\begin{equation}\label{eq:ppls_5}
\begin{aligned}
p(\vec{z}) = \mathcal{N}(\vec{0},\vec{I}).
\end{aligned}
\end{equation}

Using the likelihood $p_{\Phi}(\vec{y},\vec{s} \vert \vec{z}) = p_\upsilon(\vec{y}\vert\vec{z}) p_\zeta(\vec{s}\vert\vec{z})$, the PPLS model byperparameters~\mbox{$\vec{\Phi} = \vec{\zeta} \cup \vec{\upsilon}$} can be determined by maximising the marginal likelihood   
\begin{equation}\label{eq:ppls_6}
\begin{aligned}
p_{\Phi}(\vec{y},\vec{s}) = \int p_{\Phi}(\vec{y},\vec{s} \vert \vec{z}) p(\vec{z}) \, \D \vec{z} := \mathcal{N}(\vec{0},\hat{\vec{\Sigma}}_{ys}),
\end{aligned}
\end{equation}
which is a zero-mean Gaussian with the covariance 
\begin{equation}\label{eq:ppls_7}
\begin{aligned}
\hat{\vec{\Sigma}}_{ys} = \Biggl(\def\arraystretch{1.25}\begin{array}{c|c}
     \vec{Q}\vec{Q}^\trans + \hat{\vec{\Sigma}}_y & \vec{Q}\vec{W}^\trans\\
     \hline
     \vec{W}\vec{Q}^\trans & \vec{W}\vec{W}^\trans + \hat{\vec{\Sigma}}_s
    \end{array}
    \Biggr); 
\end{aligned}
\end{equation}
see the derivation in Appendix \hyperlink{Appendix B}{B}. 
The posterior probability density of the unobserved latent variables is given by Bayes' rule
\begin{equation}\label{eq:ppls_8}
\begin{aligned}
p_{\Phi}(\vec{z} \vert \vec{y},\vec{s}) =  \frac{p_{\Phi}(\vec{y},\vec{s}\vert\vec{z}) p(\vec{z})}{p_{\Phi}\left(\vec{y},\vec{s}\right)},
\end{aligned}
\end{equation}
which takes the form of a Gaussian probability density
\begin{equation}\label{eq:ppls_9}
\begin{aligned}
p_{\Phi}(\vec{z}\vert\vec{y},\vec{s}) := \mathcal{N}(\hat{\vec{\mu}}_{z},\hat{\vec{\Sigma}}_{z}),
\end{aligned}
\end{equation}
where the mean and covariance are given by
\begin{subequations}\label{eq:ppls_10}
\begin{align}
\hat{\vec{\mu}}_{z} &= \vec{B}^{\trans} \hat{\vec{\Sigma}}_{ys}^{-1}\vec{d},\label{eq:ppls_10a}\\
\hat{\vec{\Sigma}}_{z} &= \vec{I} - \vec{B}^{\trans}\hat{\vec{\Sigma}}_{ys}^{-1}\vec{B},\label{eq:ppls_10b}
\end{align}
\end{subequations}
and
\begin{equation}\label{eq:ppls_11}
\begin{aligned}
\vec{B} := \left(
    \begin{array}{c}
      \vec{Q}\\
      \vec{W}
    \end{array}
    \right), \quad \quad \vec{d} := \left(
    \begin{array}{c}
      \vec{y}\\
      \vec{s}
    \end{array}
    \right).
\end{aligned}
\end{equation}
The posterior probability density still depends on unknown PPLS model hyperparameters $\vec{\Phi}$. Although these could be obtained directly by maximising the marginal likelihood $p_{\Phi}(\vec{y},\vec{s})$ given in \eqref{eq:ppls_6} using gradient descent, the computational complexity is proportional to the input dimension~$d_s$. To improve computational tractability, we use variational Bayes to 
approximate the posterior with a Gaussian trial density $q(\vec{z})$ where the Kullback-Leibler (KL) divergence between the two densities is defined as
\begin{equation}\label{eq:ppls_13}
\begin{aligned}
D_{KL}\left(q(\vec{z}) \, \vert \vert \, p_{\Phi}(\vec{z}\vert\vec{y},\vec{s})\right) &= \int q(\vec{z}) \ln \left(\frac{q(\vec{z})}{p_{\Phi}(\vec{z}\vert\vec{y},\vec{s})}\right)\,\D \vec{z}\\
&= \int  q(\vec{z}) \ln \left(\frac{q(\vec{z})}{p_{\Phi}(\vec{y},\vec{s}\vert\vec{z}) p(\vec{z})}\right)\, d\vec{z} + \ln p_\Phi(\vec{y},\vec{s}).
\end{aligned}
\end{equation}
Since the KL divergence is non-negative $D_{KL}\left(\cdot \vert\vert \cdot\right) \geq 0$, the marginal likelihood is upper bounded by
\begin{equation}\label{eq:ppls_14}
\begin{aligned}
\ln p_\Phi(\vec{y},\vec{s}) 
&\geq \int q(\vec{z}) \ln \left(\frac{p_{\Phi}(\vec{y},\vec{s}\vert\vec{z}) p(\vec{z})}{q(\vec{z})}\right)\,\D \vec{z} := \mathcal{F}(\vec{y},\vec{s}).
\end{aligned}
\end{equation}
Hence, the KL divergence is minimised by maximising the evidence lower bound (ELBO) $\mathcal{F}(\vec{y},\vec{s})$, which may be written more compactly as
\begin{equation}\label{eq:ppls_15}
\begin{aligned}
\mathcal{F}(\vec{y},\vec{s}) = \expect_{q(\vec{z})}\left( \ln p_{\Phi}(\vec{y},\vec{s} \vert \vec{z})\right) + \expect_{q(\vec{z})}\left( \ln p(\vec{z})\right) - \expect_{q(\vec{z})}\left( \ln q(\vec{z})\right).
\end{aligned}
\end{equation}
Both the trial density $q(\vec{z})$ and PPLS model hyperparameters $\vec{\Phi}$ are obtained by maximising the ELBO using alternating coordinate descent with the expectation maximisation (EM) algorithm~\cite{moon1996expectation} (see Appendix \hyperlink{Appendix C}{C}). Alternatively, the PPLS model hyperparameters may also be obtained using stochastic gradient descent. 

For design optimisation problems requiring FE models with tens of thousands of nodes $n_u$, the computational time complexity can be as high as $\mathcal{O}(n_u^2)$~\cite{abdulle2008finite}. In contrast, the time complexity of EM is only approximately~$\mathcal{O}(n d_z d_s n_t + n d_z d_y n_t)$ (assuming $d_z \ll d_s \ll n$), where $n_t$ is the number of EM iterations. The latent variable dimension $d_z$ can be selected using cross-validation~\cite{hastie2009elements}. The smallest dimension that corresponds to the change in ELBO falling below a threshold is chosen. Bayesian information criterion (BIC) is a suitable alternative method for selecting the latent variable dimension~\cite{schwarz1978estimating}.\\

\subsection{Bayesian optimisation}\label{sec:bo}

BO is composed of two components, a GP surrogate to emulate the QOI and an adaptive sampling algorithm to update the GP while balancing design space exploration with exploitation of belief in where the global minimum is located.  In the proposed approach, the GP is constructed over the low-dimensional latent space.

\subsubsection{Gaussian process regression}\label{sec:gp}

\begin{figure}[b!]
	\centering
	{\includegraphics[width=80mm]{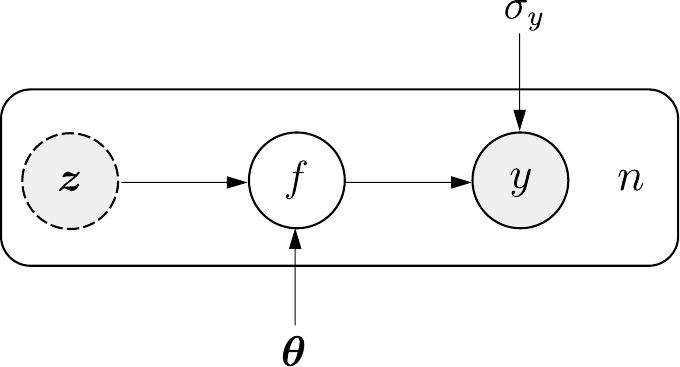}}
	\caption{Graphical model of GP regression where $n$ is the size of the training data set $\mathcal{D}$, $\vec{z}$ are the latent variables (GP inputs), $f$ is the target output variable, $y$ is the observation, and $\vec{\theta} \cup \{\sigma_y\}$ are the model hyperparameters.}
	\label{fig:gp_graphical_model}
\end{figure}

A single observation of the objective $J(\vec{W}\vec{z})$ or constraint $H^{(j)}(\vec{W}\vec{z})$ functions generically denoted by $y \in \mathbb{R}$, obeys the statistical data generating model
\begin{equation}\label{eq:gp_1}
\begin{aligned}
y = f(\vec{z}) + \epsilon_y , \quad \quad \epsilon_y \sim \mathcal{N}(0,\sigma_y^2),
\end{aligned}
\end{equation}
where $\sigma_y \in \mathbb{R}^+$ denotes the noise or error standard deviation (such as the FE modelling or discretisation error), see also Figure~\ref{fig:gp_graphical_model}. The GP $f:\mathbb{R}^{d_z}\rightarrow\mathbb{R}$ maps the latent variables $\vec{z}$ to the observation $y$ and has the prior probability density
\begin{equation}\label{eq:gp_2}
\begin{aligned}
f(\vec{z}) \sim \mathcal{GP}\left(0,c(\vec{z},\vec{z}')\right),
\end{aligned}
\end{equation}
Without loss of generality, we choose the squared-exponential covariance function $c:\mathbb{R}^{d_z} \times \mathbb{R}^{d_z} \rightarrow \mathbb{R}^+$ given by
\begin{equation}\label{eq:gp3}
\begin{aligned}
c(\vec{z},\vec{z}') = \sigma_f^2\exp\left(-\sum_{i=1}^{d_z}\frac{(z_i-z_i')^2}{2 \ell_i}\right),
\end{aligned}
\end{equation}
where $\sigma_f \in \mathbb{R}^+$ is a scaling parameter, $\ell_i \in \mathbb{R}^+$ is a length scale parameter,  and $z_i$ denotes the $i$\textsuperscript{th} component of latent variable vector $\vec{z}$. The hyperparameters of the GP are collected in the set $\vec{\theta} = \{\sigma_f,\ell_1,\ell_2,...,\ell_{d_z}\}$.

Let $\mathcal{D} = \{(\vec{z}_i,y_i)\,\vert\, i = 1,2,\cdots,n\}$ denote the training data, which is collected into the latent variable matrix \mbox{$\vec{Z} = \left(\vec{z}_1\,\vec{z}_2\,\cdots\,\vec{z}_n\right)^\trans \in \mathbb{R}^{n \times d_z}$} and output vector $\vec{y} = \left(y_1\,y_2\,\cdots\,y_n\right)^\trans \in \mathbb{R}^n$. For any test output variables $\vec{f}_* \in \mathbb{R}^{n_*}$ corresponding to test latent variables $\vec{Z}_* \in \mathbb{R}^{n_* \times d_z}$, the joint probability density with the outputs is given by
\begin{equation}\label{eq:gp_4}
\begin{aligned}
p_\Theta(\vec{f}_*,\vec{y})
=\mathcal{N}\begin{pmatrix}\begin{pmatrix}
\vec{0}\\
\vec{0}
\end{pmatrix},
\begin{pmatrix}
\vec{C}_{Z_{\!*}Z_{\!*}}  & \vec{C}_{Z_{\!*}Z}\\
\vec{C}_{ZZ_{\!*}} &  \vec{C}_{ZZ} + \sigma_y^2 \vec{I}
\end{pmatrix}
\end{pmatrix}.
\end{aligned}
\end{equation}
The entries of the covariance matrix $\vec{C}_{ZZ}$ consist of the covariance function $c(\vec{z},\vec{z}')$ evaluated at $\vec{z}$ and $\vec{z}'$, and the hyperparameters are $\vec{\Theta} = \vec{\theta} \cup \{\sigma_y\}$. Similarly, the entries of the covariance matrix $\vec{C}_{ZZ_{\!*}}$ consist of the covariance function $c(\vec{z},\vec{z}_*)$, where $\vec{z}_*$ is a row of $\vec{Z}_*$. The posterior predictive probability density is obtained by conditioning the joint probability density~$p_\Theta(\vec{f}_*,\vec{y})$ on the output vector~$\vec{y}$ and is given by 
\begin{equation}\label{eq:gp_5}
\begin{aligned}
p_\Theta(\vec{f}_*\vert\vec{y}) = \mathcal{N}\left(\vec{C}_{Z_{\!*}Z}\left(\vec{C}_{ZZ} + \sigma_y^2\vec{I}\right)^{-1}\vec{y}\,,\,\vec{C}_{Z_{\!*}Z_{\!*}} - \vec{C}_{Z_{\!*}Z}\left(\vec{C}_{ZZ} + \sigma_y^2\vec{I}\right)^{-1}\vec{C}_{ZZ_{\!*}}\right).
\end{aligned}
\end{equation}
Furthermore, the optimal GP hyperparameters $\vec{\Theta}$ can be computed by maximising the log marginal likelihood
 \begin{equation}\label{eq:gp_6}
\begin{aligned}
\vec{\Theta}^* = \argmax_\Theta \ln p_\Theta(\vec{y}),
\end{aligned}
\end{equation}
where
\begin{equation}\label{eq:gp_7}
\begin{aligned}
p_{\Theta}(\vec{y}) =  \mathcal{N}\left(\vec{0}\,,\,\vec{C}_{ZZ}+\sigma_y^2\vec{I}\right).
\end{aligned}
\end{equation}

\subsubsection{Adaptive sampling}

For any test input $\vec{z}^*$, the GP posterior probability density  $p_{\Theta_k}(f_*\vert y)$ given by \eqref{eq:gp_5} yields a predictive mean $\mu:\mathbb{R}^{d_z} \rightarrow \mathbb{R}$ and standard deviation $\sigma:\mathbb{R}^{d_z} \rightarrow \mathbb{R}^+$, which are functions of the latent variables $\vec{z}$. In the adaptive sampling iteration $k$, a new sampling point
\begin{equation}\label{eq:as_1}
\begin{aligned}
\vec{z}_{k+1} = \argmax_{\vec{z}\in D_z} \, a(\mu_k(\vec{z}),\sigma_k(\vec{z})) \,,
\end{aligned}
\end{equation}
is proposed by maximising an acquisition function $a : \mathbb{R} \rightarrow \mathbb{R}$, which balances exploration and exploitation. Although there are many kinds of acquisition functions~\cite{wilson_maximizing_2018}, upper confidence bound (UCB)~\cite{cox_statistical_1992,auer_using_2003} and expected improvement (EI)~\cite{jones_efficient_1998} are used more extensively. The UCB acquisition function
\begin{equation}\label{eq:as_2}
\begin{aligned}
a_{ucb}\left(\mu(\vec{z}),\sigma(\vec{z})\right) = -\mu(\vec{z}) + \gamma \sigma(\vec{z})\,,
\end{aligned}
\end{equation}
uses a weighted sum of the negative mean and standard deviation, where $\gamma \in \mathbb{R}^+$ is an exploration parameter. Similarly, the EI acquisition function is defined in terms of the GP mean and standard deviation and is given by
\begin{equation}\label{eq:as_3}
\begin{aligned}
a_{ei}\left(\mu(\vec{z}),\sigma(\vec{z})\right) = \left(y^+ -  \mu(\vec{z}) - \xi \right)\Phi(u) + \sigma(\vec{z})\phi(u) \,,
\end{aligned}
\end{equation}
where $\Phi: \mathbb{R} \rightarrow \mathbb{R}^+$ is the standard cumulative density function, $\phi: \mathbb{R} \rightarrow \mathbb{R}^+$ is the standard probability density function, $\xi \in \mathbb{R}$ is a jitter parameter, and
\begin{equation}\label{eq:as_4}
\begin{aligned}
u = \frac{ y^+ - \mu(\vec{z}) -  \xi}{\sigma(\vec{z})}\,.
\end{aligned}
\end{equation}

The acquisition function can be extended to prohibit solutions that do not satisfy the constraints from being proposed~\cite{gardner2014bayesian}. By denoting $f_*^{(i)} \in \mathbb{R}$ as the test output variable for the $i$\textsuperscript{th} GP corresponding to the $i$\textsuperscript{th} objective or constraint function with output vector $\vec{y}^{(i)} \in \mathbb{R}^n$, the constrained acquisition function is given by
\begin{equation}\label{eq:as_5}
\begin{aligned}
a_c\left(\mu^{(i)}(\vec{z}),\sigma^{(i)}(\vec{z})\right) = a \left (\mu^{(1)}(\vec{z}),\sigma^{(1)}(\vec{z}) \right )\prod_{i=2}^{d_y}\left(1+ \rho^{(i)} \int_{0}^\infty p_{\Theta^{(i)}}(f_*^{(i)}\vert \vec{y}^{(i)}) \, \D f_*^{(i)}\right),
\end{aligned}
\end{equation}
where $\rho^{(i)} \in \mathbb{R}$ is a penalty constant to prevent infeasible solutions being proposed. Let $\mu^{(i)}(\cdot)$ and $\sigma^{(i)}(\cdot)$ denote the mean and standard deviation of the $i$\textsuperscript{th} GP posterior predictive probability density respectively, where
\begin{equation}\label{eq:as_6}
\begin{aligned}
 p_{\Theta^{(i)}}(f_*^{(i)}\vert \vec{y}^{(i)}) = \mathcal{N}\left( \mu^{(i)}(\vec{z}_*),\left(\sigma^{(i)}(\vec{z}_*)\right)^2
 \right).
\end{aligned}
\end{equation}
Since the constrained acquisition function may be dominated by sharp peaks, genetic algorithms such as NSGA-II~\cite{deb2002fast} can be used to maximise the acquisition function. At each adaptive sampling iteration, the training data $\mathcal{D}$ is updated with the proposed sample and the GP is subsequently updated, as summarised in Algorithm \ref{alg:EGO}.

\begin{algorithm}[t!]
\caption{Adaptive sampling with BO}\label{alg:EGO}
\begin{algorithmic}
\setstretch{1.25}
\small
\State {\textbf{Data}: $\mathcal{D}_0^{(i)} = \{(\vec{z}_j,y_j^{(i)})\,\vert\,j = 1,2,\cdots , n\}$}
\State {\textbf{Input}: adaptive sampling budget $n_k$}
\For{$k \in \{0,1,\cdots,n_k-1\}$}
\State {Compute $\mu_k^{(i)}(\vec{z}_*)$ and $\sigma_k^{(i)}(\vec{z}_*)$ using $\mathcal{D}_k^{(i)}$} \Comment{posterior probability density
 \eqref{eq:gp_5}}
\State  {Propose sample $\vec{z}_{k+1} \leftarrow \argmax\limits_{\vec{z} \in D_z} a_c(\mu_k^{(i)}(\vec{z}), \sigma_k^{(i)}(\vec{z}))$} \Comment{maximise acquisition function \eqref{eq:as_5}}
\State {Compute design variables $\vec{s}_{k+1} = \vec{W}\vec{z}_{k+1}$} \Comment{linear mapping}
\State {Collect observations $y_{k+1}^{(i)}$} \Comment{solve computational model}
\State {Augment data $\mathcal{D}_{k+1}^{(i)} \leftarrow \{\mathcal{D}_k^{(i)}, (\vec{z}_{k+1},y_{k+1}^{(i)})\}$}
\EndFor
\State {\textbf{Result}: global minimum $y^{(1)*} \leftarrow \min\{y_1^{(1)},y_2^{(1)},\cdots,y_{{\scaleto{n + n_k}{5pt}}}^{(1)}\}$}
\end{algorithmic}
\end{algorithm}

\section{Multi-view Bayesian optimisation}\label{sec:adaptive_reduced}

In this section, the posterior probability density of the latent variables is computed using PPLS and embedded within the GP. We derive a posterior predictive probability density, which makes it possible to obtain empirical estimates. Finally, we summarise our approach culminating in the probabilistic partial least squares Bayesian optimisation (PPLS-BO) algorithm.

\subsection{Gaussian processes in reduced dimension}

\begin{figure*}[p!]
	\centering
	{\includegraphics[width=110mm]{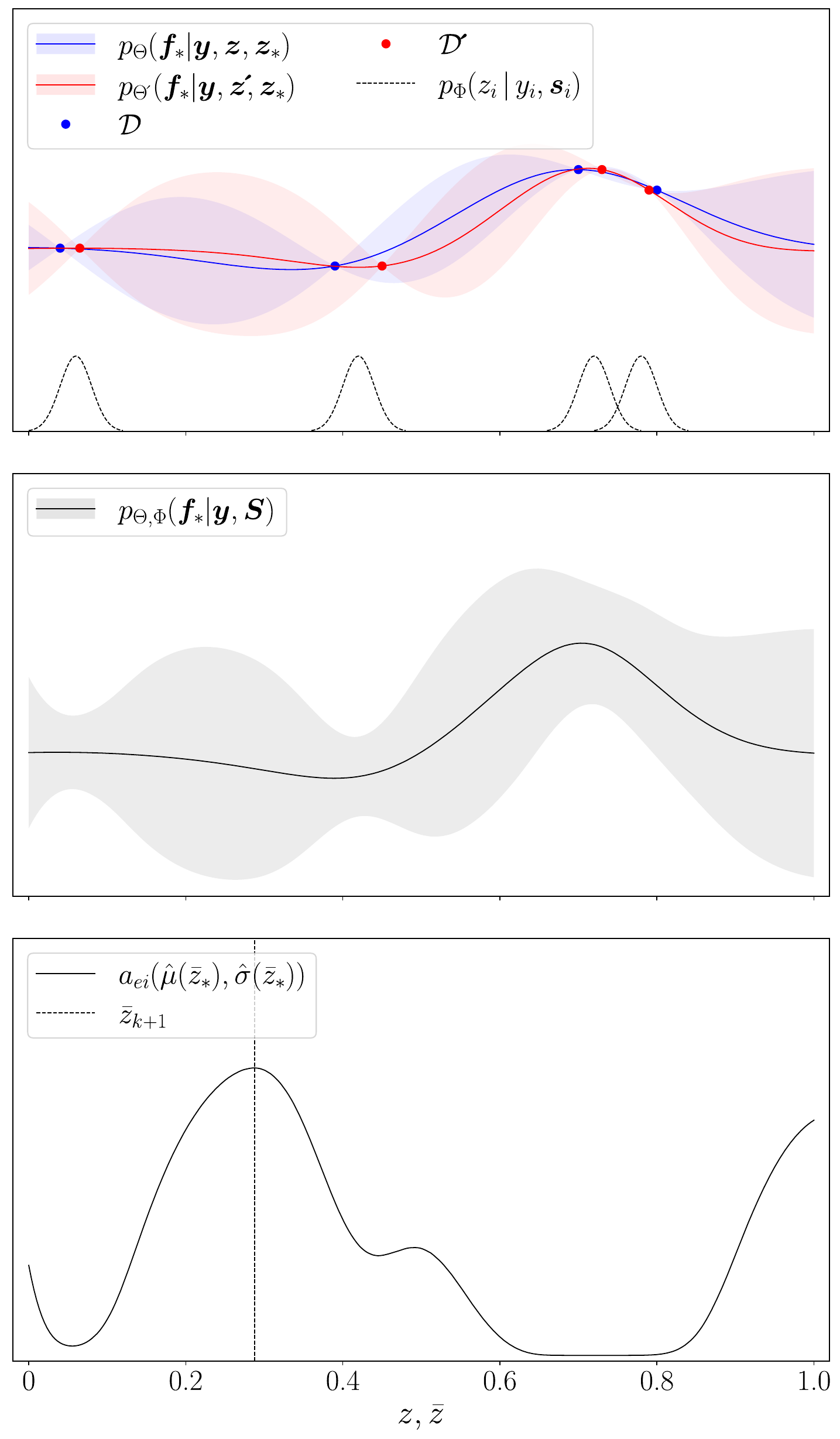}}
	\caption{Illustrative one-dimensional example for the GP regression in reduced dimension. Two realisations of the GP posterior density $p_{\Theta}(\vec{f}_*\,\vert\,\vec{y},\vec{z}, \vec{z}_*)$ and $p_{\Theta'}(\vec{f}_*\,\vert\,\vec{y},\vec{z}', \vec{z}_*)$ as in \eqref{eq:gp_5} are fitted to training data $\mathcal{D}$ and $\mathcal{D'}$ composed of the same observation vector $\vec{y}$ but different (iid) latent variables sampled from their posterior densities $\vec{z}, \vec{z}' \sim \prod_{i=1}^n p_{\Phi}(z_i\vert y_i,\vec{s}_i)$ given by~\eqref{eq:ppls-bo_2}. The marginal posterior predictive density $p_{\Theta,\Phi}\left(\vec{f}_* \vert \vec{y}, \vec{S}\right)$ from \eqref{eq:ppls-bo_1} is trained by marginalising over the training $\vec{z}$ and test $\vec{z}_*$ latent variables. In PPLS-BO, the acquisition function $a_{ei}(\cdot)$ is fitted to the mean $\hat{\mu}(\bar{\vec{z}}_*)$ and standard deviation $\hat{\sigma}(\bar{\vec{z}}_*)$ obtained from the marginal predictive posterior, and a new sample $\bar{z}_{k+1}$ is proposed.}
	\label{fig:ppls_bo_schematic}
\end{figure*}

The inputs of a standard GP, introduced in section~\ref{sec:gp}, are deterministic. However, the latent variables determined with PPLS, introduced in Section~\ref{subsec:ppls}, are random. We denote the posterior density as $p_{\Theta^{(i)}}(f_*^{(i)} \vert \vec{y}^{(i)}, \vec{Z}, \vec{z}_*)$. For each random realisation of the latent variables $\vec{Z}$ and test latent variables $\vec{z}_*$, the density $p_{\Theta^{(i)}}(f_*^{(i)} \vert \vec{y}^{(i)}, \vec{Z}, \vec{z}_*)$ may be computed as the posterior predictive probability density of a standard GP \eqref{eq:gp_5}, as shown in Figure~\ref{fig:ppls_bo_schematic}. The marginal posterior predictive probability density is obtained by marginalising over latent variables and test latent variables, that is
\begin{equation}\label{eq:ppls-bo_1}
\begin{aligned}
p_{\Theta^{(i)},\Phi}\left(f_*^{(i)}\vert \vec{Y}, \vec{S}\right) &= \int p_{\Theta^{(i)}}\left(f_*^{(i)} \vert \vec{y}^{(i)}, \vec{Z}, \vec{z}_*\right) p_\Phi(\vec{Z}\vert\vec{Y},\vec{S}) p_\Phi(\vec{z}_* \vert \vec{Y}, \vec{S}) \,\D\vec{Z} \D\vec{z}_*.
\end{aligned}
\end{equation}
The latent variables and test latent variables are given by
\begin{equation}\label{eq:ppls-bo_2}
\begin{aligned}
p_\Phi(\vec{Z}\vert\vec{Y},\vec{S}) &= \prod_{i=1}^n p_{\Phi}\left(\vec{z}_i\vert\vec{y}_i,\vec{s}_i\right),
\end{aligned}
\end{equation}
and
\begin{equation}\label{eq:ppls-bo_3}
\begin{aligned}
p_{\Phi}(\vec{z}_*\vert\vec{Y},\vec{S}) &= \mathcal{N}\left(\bar{\vec{z}}_*,\hat{\vec{\Sigma}}_z\right). 
\end{aligned}
\end{equation}
Here, we assumed independently and identically distributed samples. The design variable matrix $\vec{S} = (\vec{s}_1 \, \vec{s}_2 \cdots \vec{s}_n)^\trans \in \mathbb{R}^{n \times d_s}$ collects the $n$ design variables and output vectors $\vec{y}^{(i)}$ consisting of observations corresponding to the $i$\textsuperscript{th} objective or constraint function are collected into the matrix \mbox{$\vec{Y} = (\vec{y}^{(1)} \, \vec{y}^{(2)} \cdots \vec{y}^{(n)}) \in \mathbb{R}^{n \times d_y}$}. The latent variable posterior probability density $p_{\Phi}(\vec{z}_i\vert\vec{y}_i,\vec{s}_i)$ is computed using the PPLS model \eqref{eq:ppls_9}, and the test latent variable posterior density has the same covariance $\hat{\vec{\Sigma}}_z$ given by \eqref{eq:ppls_10b}. The mean test latent variables $\bar{\vec{z}}_*$ now become the independent variables for optimisation. Test output variables $f_*^{(i)}$ used for surrogate-based design optimisation can be sampled from the marginal posterior probability density $p_{\Theta^{(i)},\Phi}(f_*^{(i)}\vert \vec{Y}, \vec{S})$ yielding an expectation (by the law of total expectation)
\begin{equation}\label{eq:ppls-bo_4}
\begin{aligned}
\expect \left(f^{(i)}(\vec{z}_{*})\right) &= \expect_{p_{\Phi}(z_{*} \vert Y, S)}\left(\expect_{p_{\Phi}(Z \vert Y, S)}\left( \expect_{p_{\Theta^{(i)}}(f_*^{(i)} \vert y^{(i)},Z,z_*)}\left( f_*^{(i)} \right) \right)\right) \quad \quad \quad\\
&= \expect_{p_{\Phi}(z_{*} \vert Y, S)}\left(\expect_{p_{\Phi}(Z \vert Y, S)}\left( \mu^{(i)}(\vec{z}_{*}) \right)\right)\\
&= \expect\left( \mu^{(i)}(\vec{z}_{*})\right),
\end{aligned}
\end{equation}
and variance (by law of total variance)
\begin{equation}\label{eq:ppls-bo_5}
\scalebox{0.875}{$
\begin{aligned}
\var \left(f^{(i)}(\vec{z}_{*})\right) &= \var_{p_{\Phi}(z_{*} \vert Y, S)}\left(\var_{p_{\Phi}(Z \vert Y, S)}\left( \expect_{p_{\Theta^{(i)}}(f_*^{(i)} \vert y^{(i)},Z,z_*)}\left( f_*^{(i)} \right) \right)\right) +\\
&\quad \quad \,\,\,\expect_{p_{\Phi}(z_{*} \vert Y, S)}\left(\expect_{p_{\Phi}(Z \vert Y, S)}\left( \var_{p_{\Theta^{(i)}}(f_*^{(i)} \vert y^{(i)},Z,z_*)}\left( f_*^{(i)} \right)\right)\right)\\
&= \var_{p_{\Phi}(z_{*} \vert Y, S)}\left(\var_{p_{\Phi}(Z \vert Y, S)}\left(\mu^{(i)}(\vec{z}_{*}) \right)\right) +  \expect_{p_{\Phi}(z_{*} \vert Y, S)}\left(\expect_{p_{\Phi}(Z \vert Y, S)}\left( \left(\sigma^{(i)}(\vec{z}_{*}) \right)^2\right)\right)\\
&= \var\left( \mu^{(i)}(\vec{z}_{*})\right) + \expect\left( \left(\sigma^{(i)}(\vec{z}_{*})\right)^2\right).
\end{aligned}
	$}
\end{equation}
Both the expectation and variance can be estimated numerically using Monte Carlo (MC) sampling, yielding an approximate marginal posterior probability density
\begin{equation}\label{eq:ppls-bo_6}
\begin{aligned}
p_{\Theta^{(i)},\Phi}\left(f_*^{(i)}\vert \vec{Y}, \vec{S}\right) \approx \hat{p}_{\Theta^{(i)},\Phi}\left(f_*^{(i)}\vert \vec{Y}, \vec{S}\right) = \mathcal{N}\left(\hat{\mu}^{(i)}(\bar{\vec{z}}_*),\left(\hat{\sigma}^{(i)}(\bar{\vec{z}}_*)\right)^2\right),
\end{aligned}
\end{equation}
where
\begin{subequations}\label{eq:ppls-bo_7}
\begin{align}
\hat{\mu}^{(i)}(\bar{\vec{z}}_*) &= \expect\left( \mu^{(i)}(\vec{z}_{*})\right),\label{eq:ppls-bo_7a}\\
\hat{\sigma}^{(i)}(\bar{\vec{z}}_*) &= \sqrt{\var\left( \mu^{(i)}(\vec{z}_{*})\right) + \expect\left( \left(\sigma^{(i)}(\vec{z}_{*})\right)^2\right)}.\label{eq:ppls-bo_7b}
\end{align}
\end{subequations}

\subsection{PPLS-BO algorithm}

\begin{figure*}[t!]
\centering
{\includegraphics[width=130mm]{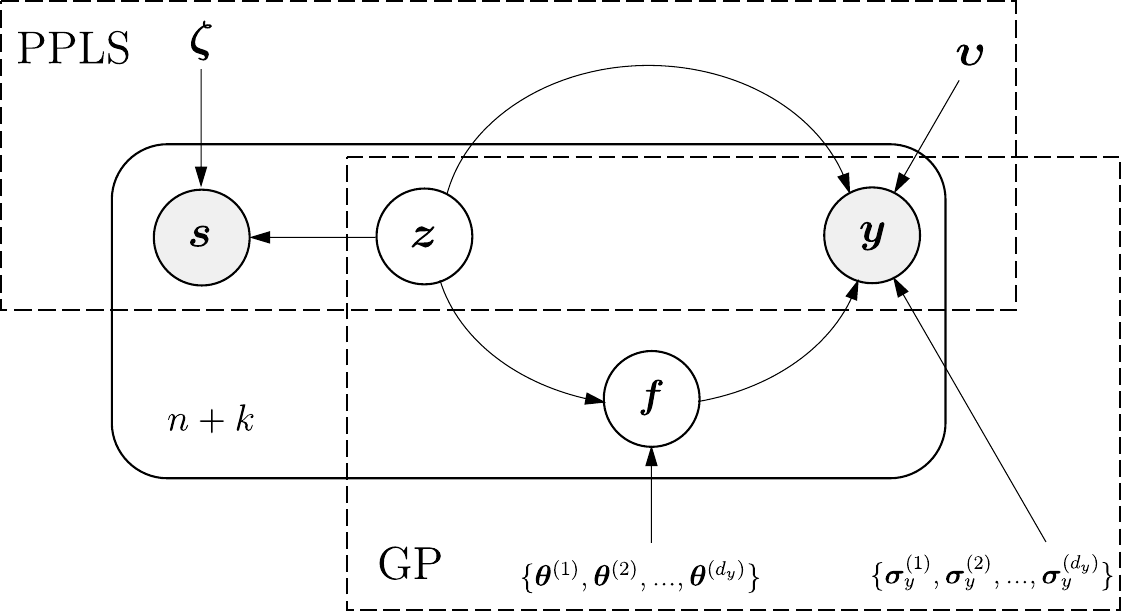}}
\caption{Graphical model for the $k$\textsuperscript{th} adaptive sampling iteration of PPLS-BO with a training data set $\mathcal{D}$ of size $n+k$ with (random) design variables $\vec{s}$ and observations \mbox{$\vec{y}=(y^{(1)}\,\,y^{(2)}\,\,...\,\,y^{(d_y)})$}. Unobserved latent variables $\vec{z}$ and hyperparameters \mbox{$\vec{\zeta} = \{\vec{W},\hat{\vec{\Sigma}}_s\}$} and $\vec{\upsilon} = \{\vec{Q},\hat{\vec{\Sigma}}_y\}$ are learned using PPLS independent from the GP surrogates with target output variables \mbox{$\vec{f}=(f^{(1)}\,\,f^{(2)}\,\,...\,\,f^{(d_y)})$} and GP hyperparameters $\{\vec{\theta}^{(1)},\vec{\theta}^{(2)},...,\vec{\theta}^{(d_y)}\}$ and $\{\sigma_y^{(1)},\sigma_y^{(2)},...,\sigma_y^{(d_y)}\}$.}
\label{fig:ppls_bo}
\end{figure*}

We present the PPLS-BO algorithm, which sequentially combines the PPLS model and GP surrogate for adaptive sampling in reduced dimension. At each adaptive sampling iteration, the PPLS model is fitted to obtain the posterior probability densities $p_\Phi(\vec{Z}\vert\vec{Y},\vec{S})$ and $p_{\Phi}(\vec{z}_*\vert\vec{Y},\vec{S})$ given by \eqref{eq:ppls-bo_2} and \eqref{eq:ppls-bo_3} respectively. For each pair of latent variables $\vec{Z}$ and test latent variables $\vec{z}_*$ drawn from their respective densities, a GP posterior predictive probability density $p_{\Theta^{(i)}}(f_*^{(i)} \vert \vec{y}^{(i)}, \vec{Z}, \vec{z}_*)$ given by \eqref{eq:gp_5} is computed. The hyperparameters of the PPLS model $\vec{\Phi}$ and each GP surrogate $\vec{\Theta}^{(i)}$ are learned independently by minimising the KL divergence using the EM algorithm (Appendix \hyperlink{Appendix C}{C}) and maximising the log marginal likelihood \eqref{eq:gp_6} respectively (Figure \ref{fig:ppls_bo}).

\begin{figure*}[b!]
	\centering
	{\includegraphics[width=100mm]{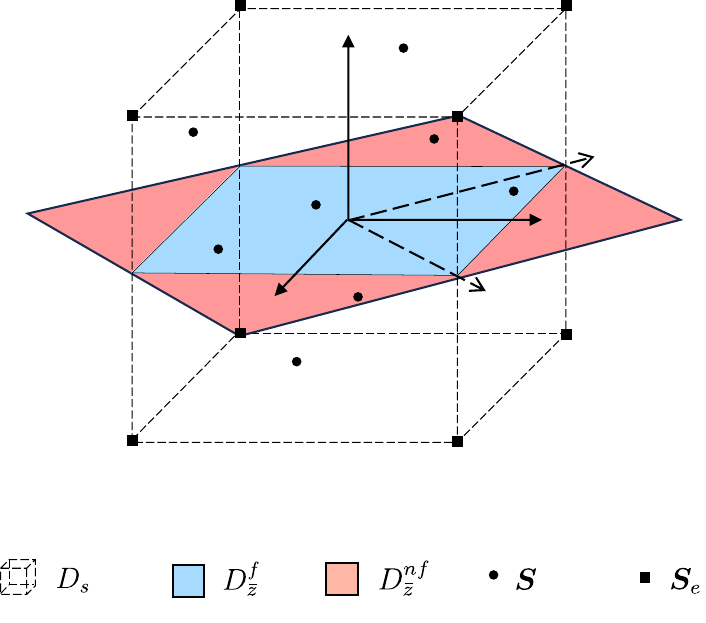}}
	\caption{Schematic of the mean latent variable domain $D_{\bar{z}}$ composed of the union between the feasible subdomain $D_{\bar{z}}^f$ contained within the design variable domain $D_s$, and the infeasible subdomain $D_{\bar{z}}^{nf}$ which extends beyond the design variable domain, where $\vec{S}$ are the design variables and $\vec{S}_e$ are the extreme design variables found at the vertices of the hyperplane describing the design variable domain.}
	\label{fig:mvl_domain}
\end{figure*}

The mean \eqref{eq:ppls-bo_4} and variance \eqref{eq:ppls-bo_5} of the marginal posterior predictive density $p_{\Theta^{(i)},\Phi}(f_*^{(i)}\vert \vec{Y}, \vec{S})$ given by \eqref{eq:ppls-bo_1}, is computed by MC sampling from the latent variable posterior densities and the GP posterior predictive density. This predictive mean and variance are inputs to the acquisition function \eqref{eq:as_1}, which is subsequently maximised to propose mean test latent variables $\bar{\vec{z}}$. The mean test latent variables exist in the bounded domain
\begin{equation}\label{eq:ppls_bo_8}
\begin{aligned}
D_{\bar{z}}^f = \left\{ \bar{\vec{z}} \in \mathbb{R}^{d_z} \, \Big\vert \, \vec{W}\bar{\vec{z}} \in D_s \right\},
\end{aligned}
\end{equation}
such that the mean of the probability density $p_\zeta(\vec{s} \vert \bar{\vec{z}})$ given by \eqref{eq:ppls_4a} is contained within the original design variable domain
$D_s$ given in \eqref{eq:ddo_1}. Since the latent variables are a linear combination of the design variables, the feasible subdomain $D_{\bar{z}}^f$ represents a convex hull. This may be contained within a minimum bounding box $D_{\bar{z}}$ that represents the union of the feasible subdomain $D_{\bar{z}}^f$ and an infeasible region $D_{\bar{z}}^{nf}$ that exists outside the original design variable domain, see Figure \ref{fig:mvl_domain}. During optimisation, infeasible solutions are prevented from being proposed by multiplying the constrained acquisition function \eqref{eq:as_5} by the indicator function

\begin{equation}\label{eq:ppls_bo_9}
\begin{aligned}
\Delta(\vec{W}\bar{\vec{z}}) =
\left\{
\begin{array}{l}
1, \quad  \text{if} \,\,\vec{W}\bar{\vec{z}} \in D_s\\
0, \quad \text{if} \,\,\vec{W}\bar{\vec{z}} \notin D_s
\end{array}
\right.
\end{aligned}.
\end{equation}

Each mean latent variable $\bar{\vec{z}}$ proposed at each adaptive sampling iteration is mapped to a corresponding conditional density~$p_\zeta(\vec{s} \vert \bar{\vec{z}})$ from which design variables $\vec{s}$ are sampled. The conditional density has a diagonal covariance matrix $\hat{\vec{\Sigma}}_s$ with each entry being the variance, which represents the error in reconstructing each design variable in the vector $\vec{s}$ from the low-rank approximation $\vec{s} \approx \vec{W}\bar{\vec{z}}$. Entries of the covariance matrix $\hat{\vec{\Sigma}}_s$ become small when the linear subspace basis $\vec{W}$ is aligned with the axes of the original design space, i.e. where the reconstruction error is low. Conversely, the entries of the covariance matrix become large when the reconstruction error is large. Instead of neglecting exploration along axes of the original design space misaligned with the linear subspace basis, we sample randomly along these axes. Where the latent variable dimension $d_z$ is specified incorrectly (for example, is too small), locations that do not exist within the linear subspace are still explored by sampling randomly from the conditional density $p_\zeta(\vec{s} \vert \bar{\vec{z}})$. For accurate low-rank approximations, sampling from the conditional density does not adversely affect the ability for adaptive sampling to converge towards the global minimum.

For each design variable vector $\vec{s}$ proposed at each adaptive sampling iteration, the expensive-to-evaluate computational model (for example, a PDE) is solved to obtain an output vector $\vec{y}$ consisting of objective and constraint function evaluations pertaining to each QOI. Design variable $\vec{S}$ and output $\vec{Y}$ matrices are augmented with the proposed design variable vector $\vec{s}$ and output vector $\vec{y}$ respectively. The adaptive sampling iterations repeat until a computational budget $n_k$ is exceeded or convergence criteria is satisfied, as summarised in Algorithm \ref{alg:PPLS-BO}. 

\begin{algorithm}[b!]
\caption{PPLS-BO}\label{alg:PPLS-BO}
\begin{algorithmic}
\setstretch{1.25}
\small
\State {\textbf{Data}: $\vec{S},\vec{Y}$}\Comment{mean centred and normalised}
\State {\textbf{Input}: adaptive sampling budget $n_k$, EM budget $n_t$, MC sampling budget $n_l$,\\
$\quad\quad\,\,\,\,$ linear subspace dimension $d_z$}
\State {Initialise $\vec{S}_0 \leftarrow \vec{S},\vec{Y}_0 \leftarrow \vec{Y}$}
\For{$k \in \{0,1,\cdots,n_k-1\}$}
\State {Fit PPLS model $\vec{\Phi}_k \leftarrow \mathrm{EM}(\vec{S}_k,\vec{Y}_k,n_{t},d_z)$}\Comment{Algorithm \ref{alg:em}}
\State {Compute $p_{\Phi_k}\left(\vec{Z}_k \vert \vec{Y}_k,\vec{S}_k\right)$ and $p_{\Phi_k}(\vec{z}_* \vert \vec{Y}, \vec{S})$}\Comment{ posterior densities \eqref{eq:ppls-bo_2},\eqref{eq:ppls-bo_3}}
\For{$l \in \{0,1,\cdots,n_l-1\}$}\Comment{MC sampling}
\State{Sample $\vec{Z}_k \sim p_{\Phi_k}\left(\vec{Z}_k \vert \vec{Y}_k,\vec{S}_k\right)$ and $\vec{z}_* \sim p_{\Phi_k}(\vec{z}_* \vert \vec{Y}_k, \vec{S}_k)$}
\State{Sample $ \left(f_*^{(i)}\right)_l \sim p_{\Theta_k^{(i)}}\left(f_*^{(i)} \vert \vec{y}_k^{(i)}, \vec{Z}_k, \vec{z}_*\right)$}\Comment{GP posterior density \eqref{eq:gp_5}}
\EndFor
\State {Compute $\hat{\mu}^{(i)}(\bar{\vec{z}}_*)$ and $\hat{\sigma}^{(i)}(\bar{\vec{z}}_*)$ using samples \scalebox{0.9}{$\{(f_*^{(i)})_1,(f_*^{(i)})_2,\cdots,(f_*^{(i)})_{n_l}\}$}}\Comment{marginal density \eqref{eq:ppls-bo_6}}
\State {Define latent variable domain $D_{\bar{z}}^f$} \Comment{approximated using \eqref{eq:ppls_bo_8}}
\State  {Propose latent variable \scalebox{0.9}{$\bar{\vec{z}}_{k+1} \leftarrow \argmax\limits_{\bar{z} \in D_z^f} a_{c}\left(\hat{\mu}_k^{(i)}(\bar{\vec{z}}), \hat{\sigma}_k^{(i)}(\bar{\vec{z}})\right)\Delta(\vec{W}_k\bar{\vec{z}})$}}\Comment{acquisition function \eqref{eq:as_5},\eqref{eq:ppls_bo_9}}
\State {Sample design parameter $\vec{s}_{k+1} \sim p_{\zeta_k}(\vec{s}_{k+1}\vert \bar{\vec{z}}_{k+1})$} \Comment{posterior density \eqref{eq:ppls_4a}}
\State {Solve for observation $y_{k+1}^{(i)}$, $\vec{y}_{k+1}\leftarrow\left(y_{k+1}^{(1)},y_{k+1}^{(2)},\cdots,y_{k+1}^{(d_y)}\right)^\trans$}
\State {Augment data \scalebox{0.85}{$\vec{S}_{k+1} \leftarrow \left(\vec{S}_k^\trans,\vec{s}_{k+1}\right)^\trans$, $\vec{Y}_{k+1} \leftarrow \left(\vec{Y}_k^\trans,\vec{y}_{k+1}\right)^\trans$}}
\EndFor
\State {\textbf{Result}: global minimum $\left(y^{(1)}\right)^* \leftarrow \min \left\{y_1^{(1)},y_2^{(1)},\cdots,y_{n + n_k}^{(1)}\right\}$}
\end{algorithmic}
\end{algorithm}

\section{Examples}\label{sec:examples}

We demonstrate the computational advantages of the proposed PPLS-BO algorithm on three numerical examples of increasing complexity. We compare our PPLS-BO algorithm with its deterministic counterpart PLS-BO, see~\cite{amine2018efficient} and Appendix \hyperlink{Appendix A}{A}, and BO in reduced dimension using a latent space discovered using PCA (PCA-BO), and classical BO. To begin, an illustrative example visually demonstrates the robustness of our approach to latent variable dimension misspecification. Next, a cantilever beam example shows the computational advantages of using PPLS-BO. Finally, we demonstrate the use of PPLS-BO on a complex manufacturing example that is otherwise difficult to optimise using classical methods. In all examples, the acquisition function is maximised using genetic algorithms (i.e. NSGA-II) with tuned optimisation parameters. 

\subsection{Illustrative example}

For the design optimisation problem \eqref{eq:ddo_1}, we consider the multi-modal objective function given by
\begin{equation}\label{eq:ex1_1}
\begin{aligned}
J\left(\vec{s}\right) = \frac{1}{9}\left(6s_1^2 + 3\right)\sin{\left(9s_1^2+1\right)}\cos{\left(6s_2^2+2\right)} + \frac{1}{1000}\sum_{i=3}^{20}s_i,
\end{aligned}
\end{equation}
with the constraint
\begin{equation}\label{eq:ex1_2}
\begin{aligned}
H\left(\vec{s}\right) = \frac{3}{4} - s_1 - s_2 - \frac{1}{1000}\sum_{i=3}^{20}s_i,
\end{aligned}
\end{equation}
which are both functions of 20 design variables $\vec{s}$ ($d_s = 20$) with an intrinsic dimension of two, i.e. $d_z = 2$. The goal is to compute the optimum design variables $\vec{s}^*$ that minimise the objective $J(\vec{s})$ subject to the constraint $H(\vec{s}) \leq 0$. The design variables are sampled from the domain $D_s = \{ \vec{s} \in \mathbb{R}^{20} \, \vert \, 0 \leq s_i \leq 1, \, i = 1,2,\cdots,20\}$. The training data set $\mathcal{D}$ is initialised with $n=24$ samples, comparing samples initialised randomly using Latin hypercube sampling (LHS) with orthogonal samples generated using a Plackett–Burman design (PBD). For a two-level PBD, low and high levels of 0 and 1 for each design variable are sampled, respectively.

For training data initialised using PBD, both PPLS-BO and PLS-BO correctly discover the intrinsic dimensionality, as shown by the large components in the first and second rows of the initial basis $\vec{W}_0$ demonstrating that $s_1$ and $s_2$ are the most influential design variables (Table \ref{tab:ex1_init}). However for PCA-BO which discovers the latent space using PCA, latent variables $\vec{z}$ are not informed by observations of the objective function $J(\vec{s})$ and consequently, the intrinsic dimensionality is not fully described by the initial basis $\vec{W}_0$, as shown by the relatively small components in the first and second rows. Consequently to apply PCA-BO effectively, the latent space dimension $d_z$ would have to be increased, to permit further exploration during adaptive sampling. However, when initialising the training data using LHS, both PPLS-BO and PLS-BO algorithms are also unable to correctly discover the intrinsic dimensionality. This outcome was observed when using both plain and orthogonal LHS to initialise the training data. Consequently, this incurs a larger reconstruction error estimate in the PPLS model. Since LHS samples are not axially-aligned or stratified, confounding between the effects of each design variable on the objective and constraint function variance is observed. Using the basis $\vec{W}_0$ computed from LHS samples would restrict exploration in the domain of the most influential design variables $s_1$ and $s_2$. Consequently, we use orthogonal samples generated from a PBD.

To test the robustness of the PPLS-BO algorithm, we consider the case where the latent variables are misspecified to be of a single dimension ($d_z=1$) and compare to PLS-BO. The optimum design variables $\vec{s}^*$ are computed using the constrained acquisition function \eqref{eq:as_5} paired with the EI acquisition function \eqref{eq:as_3} (with jitter parameter $\xi=0$) fitted to the objective function GP. The training data set $\mathcal{D}$ is initialised using a PBD, with three additional space-filling data points sampled using LHS (to initialise the GP surrogate). An adaptive sampling budget of $n_k=10$ is used, with an EM budget $n_t=100$ and MC sampling budget $n_l=10^3$.

\begin{figure*}[h!]
\centering
\subfloat[Iteration $k=0$ \label{fig:ex1_1d_a}]{\includegraphics[width=138mm]{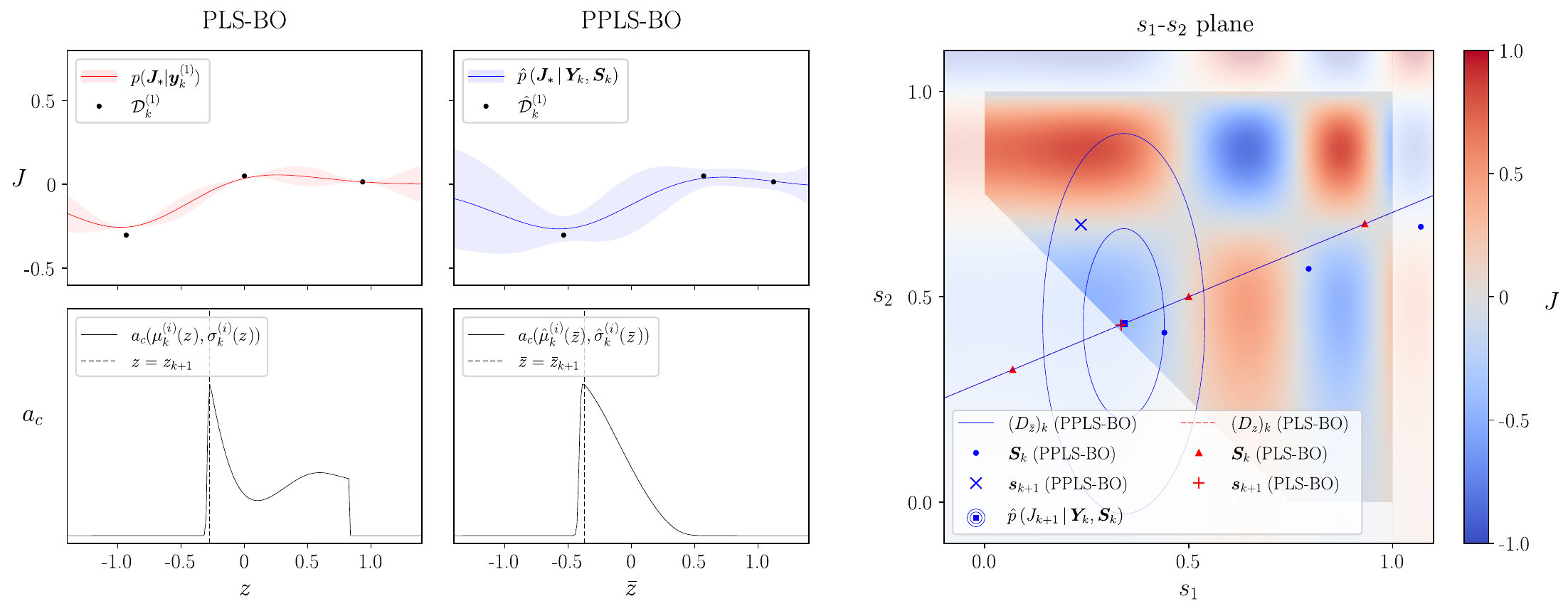}}\\
\subfloat[Iteration $k=2$ \label{fig:ex1_1d_b}]{\includegraphics[width=138mm]{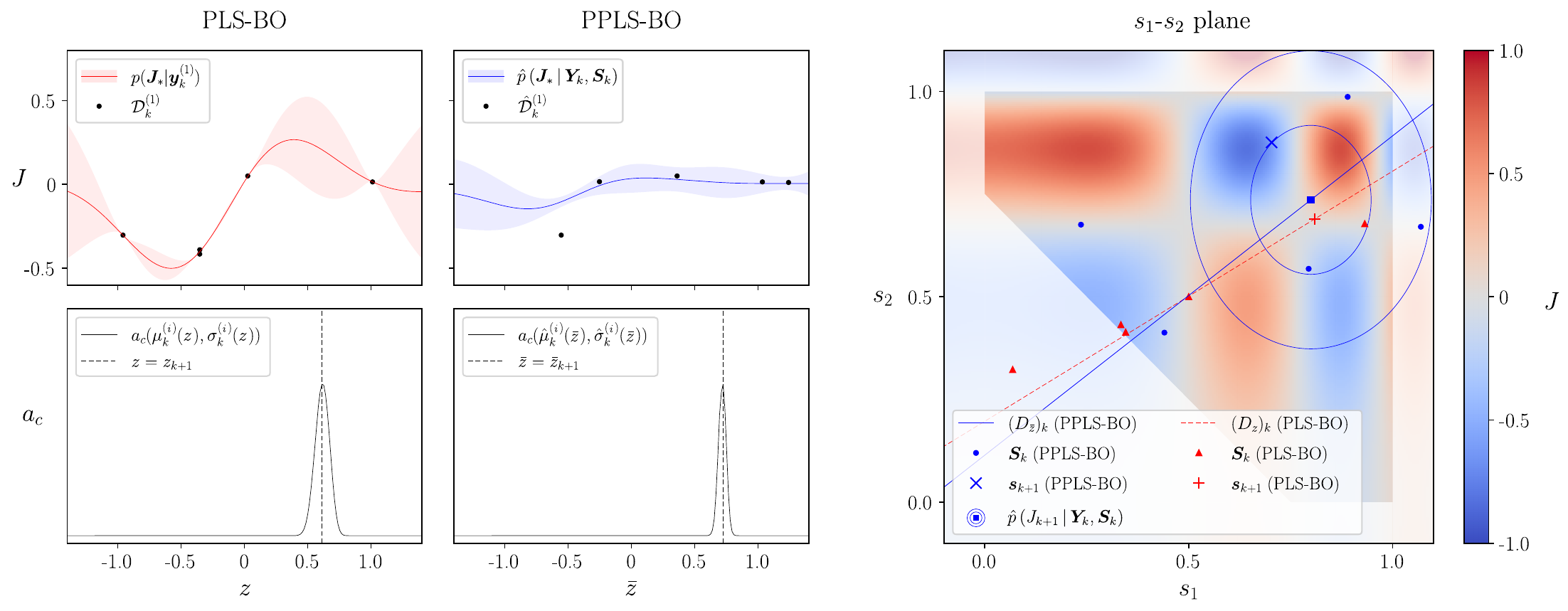}}\\
\subfloat[Iteration $k=8$ \label{fig:ex1_1d_c}]{\includegraphics[width=138mm]{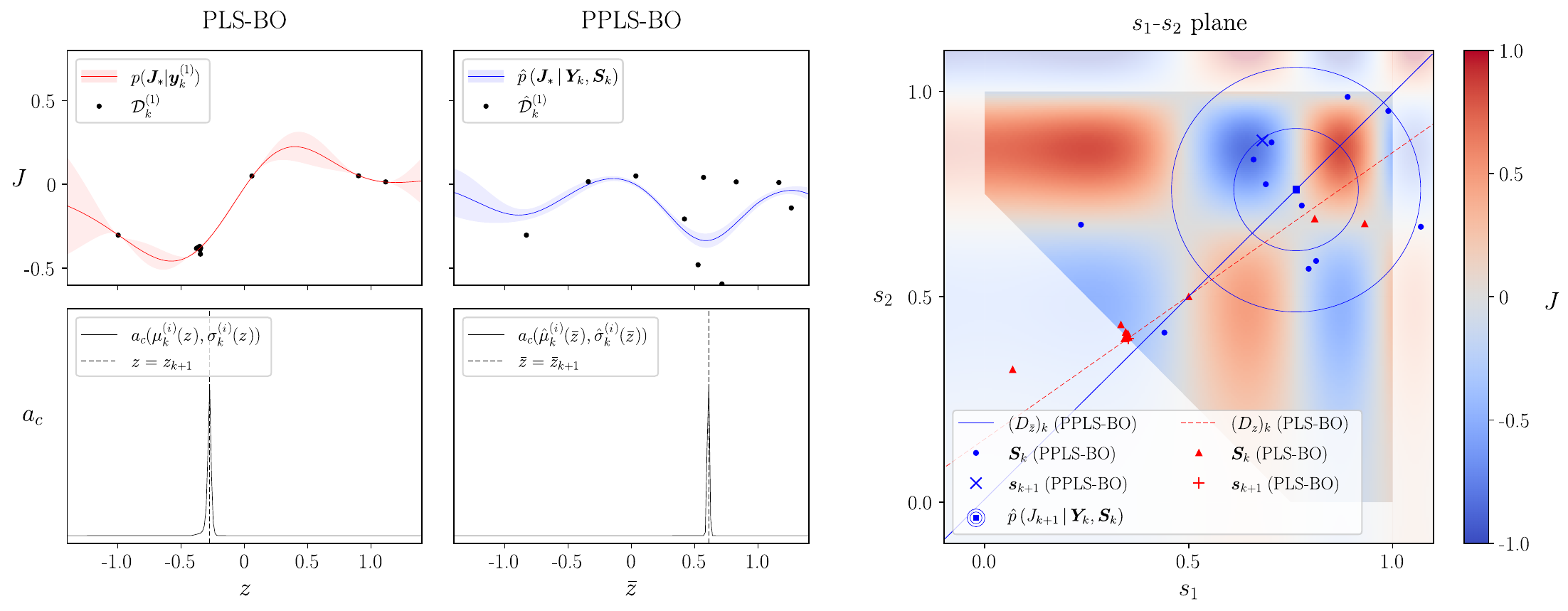}}
\caption{Illustrative example. Solution field of the objective function $J(\vec{s})$ showing the PLS-BO posterior predictive density $p(\vec{J}_* \vert \vec{y}_k^{(1)})$ fitted to data \mbox{$\mathcal{D}_k^{(1)} = \{(z_i,y_i^{(1)}) \, \vert \, i = 1,2,\cdots,n + k \}$} and acquisition function $a_c(\mu_k^{(i)}(z),\sigma_k^{(i)}(z))$ with proposed sample $z_{k+1}$. This is compared to the PPLS-BO marginal posterior predictive density $\hat{p}(\vec{J}_* \vert \vec{Y}_k,\vec{S}_k)$ showing the mean training data points \mbox{$\hat{\mathcal{D}}_k^{(1)} = \{((\hat{\mu}_z)_i,y_i^{(1)}) \, \vert \, i = 1,2,\cdots,n + k\}$} with acquisition function $a_c(\hat{\mu}_k^{(i)}(\bar{z}),\hat{\sigma}_k^{(i)}(\bar{z}))$ and proposed sample $\bar{z}_{k+1}$. Both models are projected onto the $s_1s_2$ plane showing the proposed design variable $\vec{s}_{k+1}$ and corresponding marginal posterior probability density $\hat{p}(J_{k+1} \vert \vec{Y}_k,\vec{S}_k)$ for adaptive sampling iterations (a) $k=0$, (b) $k=2$, and (c) $k=8$.}
\label{fig:ex1_1d}
\end{figure*}

Initially (at adaptive sampling iteration $k=0$), the basis $\vec{W}_k$ of the linear subspace computed using PLS-BO and PPLS-BO are approximately identical (Figure \ref{fig:ex1_1d_a}). However, the GP marginal posterior probability density $\hat{p}(\vec{J}_* \, \vert \, \vec{Y}_k, \vec{S}_k)$ computed using PPLS-BO has a similar mean, but a significantly larger variance when compared to the GP posterior probability density $p(\vec{J}_* \, \vert \, \vec{y}_k^{(1)})$ fitted using PLS-BO. This is due to the latent variable $\vec{z}$ being uncertain in PPLS-BO, which increases the total uncertainty when inferring the objective function $J(\vec{s})$ \eqref{eq:ex1_1}. The proposed posterior probability density $\hat{p}(J_{k+1}\vert \vec{Y}_k,\vec{S}_k)$ shows small and large variance in the prediction of the first $s_1$ and second $s_2$ design variables respectively, with this being correlated to the basis vectors (columns of $\vec{W}_k$). 

At the third adaptive sampling iteration ($k=2$), the basis $\vec{W}_k$ slightly differs between the PPLS-BO and PLS-BO algorithms. The GP marginal posterior probability density $\hat{p}(\vec{J}_* \, \vert \, \vec{Y}_k, \vec{S}_k)$ computed using PPLS-BO differs substantially from the GP posterior probability density $p(\vec{J}_* \, \vert \, \vec{y}_k^{(1)})$ fitted using PLS-BO, both in the mean and variance prediction (Figure \ref{fig:ex1_1d_b}). However, samples $\bar{z}_{k+1}$ (PPLS-BO) and $z_{k+1}$ (PLS-BO) are proposed at approximately the same location. Due to the change in basis, the variance has both increased and decreased in the first $s_1$ and second $s_2$ design variables respectively, when compared to the initialised models (at adaptive sampling iteration $k=0$). Contrary to PLS-BO, PPLS-BO proposes a probability density $\hat{p}(J_{k+1}\vert \vec{Y}_k,\vec{S}_k)$ which estimates reconstruction error and permits further exploration. Consequently, the design variables $\vec{s}_{k+1}$ proposed using the PPLS-BO algorithm are close to the global minimum of~\mbox{$J(\vec{s}^*)=-0.817$} at \mbox{$\vec{s}^* = (0.642\,\,0.858\,\,0\,\,0\cdots0)^\trans$}, a location that would otherwise not be proposed if using the deterministic counterpart PLS-BO.

PPLS-BO and PLS-BO algorithms both converge to different solutions, as demonstrated at the ninth adaptive sampling iteration ($k=8$) (Figure~\ref{fig:ex1_1d_c}). A non-probabilistic treatment confines exploration to the latent space, as shown by solutions only being proposed on the red line in Figure~\ref{fig:ex1_1d}. Consequently, PLS-BO cannot locate the global minimum. The fully Bayesian treatment extends the search outside the latent space by randomly sampling the remaining dimensions, as shown by solutions being proposed from a probability density centred on the blue line in Figure~\ref{fig:ex1_1d}. The posterior probability densities computed using the PPLS-BO and PLS-BO algorithms differ substantially. For PPLS-BO, similar values of the mean latent variable $\bar{z}$ yield substantially different values of the objective function $J(\vec{s})$, resulting in a large estimate of the observation standard deviation, where $\sigma_y=0.153$. Conversely, the latent variable $z$ is deterministic in the PLS-BO model, resulting in a minimal observation standard deviation, where $\sigma_y=0.002$. 

\begin{table}[t!]
\begin{center}
\caption{The linear subspace basis $\vec{W}_0$ for the illustrative example in the first adaptive sampling iteration for LHS and PBD data initialisation methods using the PCA-BO, PLS-BO, and PPLS-BO algorithms.}
\scalebox{0.975}{$
\begin{tabular}{cccc}
\toprule
Data Initialisation & PCA-BO & PLS-BO & PPLS-BO \\ \cmidrule{1-4}
LHS &$\left(\begin{array}{cc} 
	-0.027 & 0.427 \\
	0.331 & 0.086 \\
	\vdots &  \vdots
\end{array}\right)$  & $\left(\begin{array}{cc} 
    0.567 & 0.021 \\
    0.518 & 0.294 \\
    \vdots &  \vdots
    \end{array}\right)$ &   $\left(\begin{array}{cc} 
    0.564 & 0.093\\
    0.516 & 0.302\\
    \vdots &  \vdots
    \end{array}\right)$\vspace{5pt} \\
PBD & $\left(\begin{array}{cc} 
	0.208 & -0.259 \\
	0.050 & 0.337 \\
	\vdots &  \vdots
\end{array}\right)$ & $\left(\begin{array}{cc} 
    0.925 & -0.380 \\
    0.381 & 0.925 \\
    \vdots &  \vdots
    \end{array}\right)$ &   $\left(\begin{array}{cc} 
    0.925 & -0.379 \\
    0.381 & 0.926 \\
    \vdots &  \vdots
    \end{array}\right)$        \\\bottomrule
\end{tabular}
$}
\label{tab:ex1_init}
\end{center}
\end{table}

\begin{figure*}[b!]
	\centering
	\includegraphics[width=100mm]{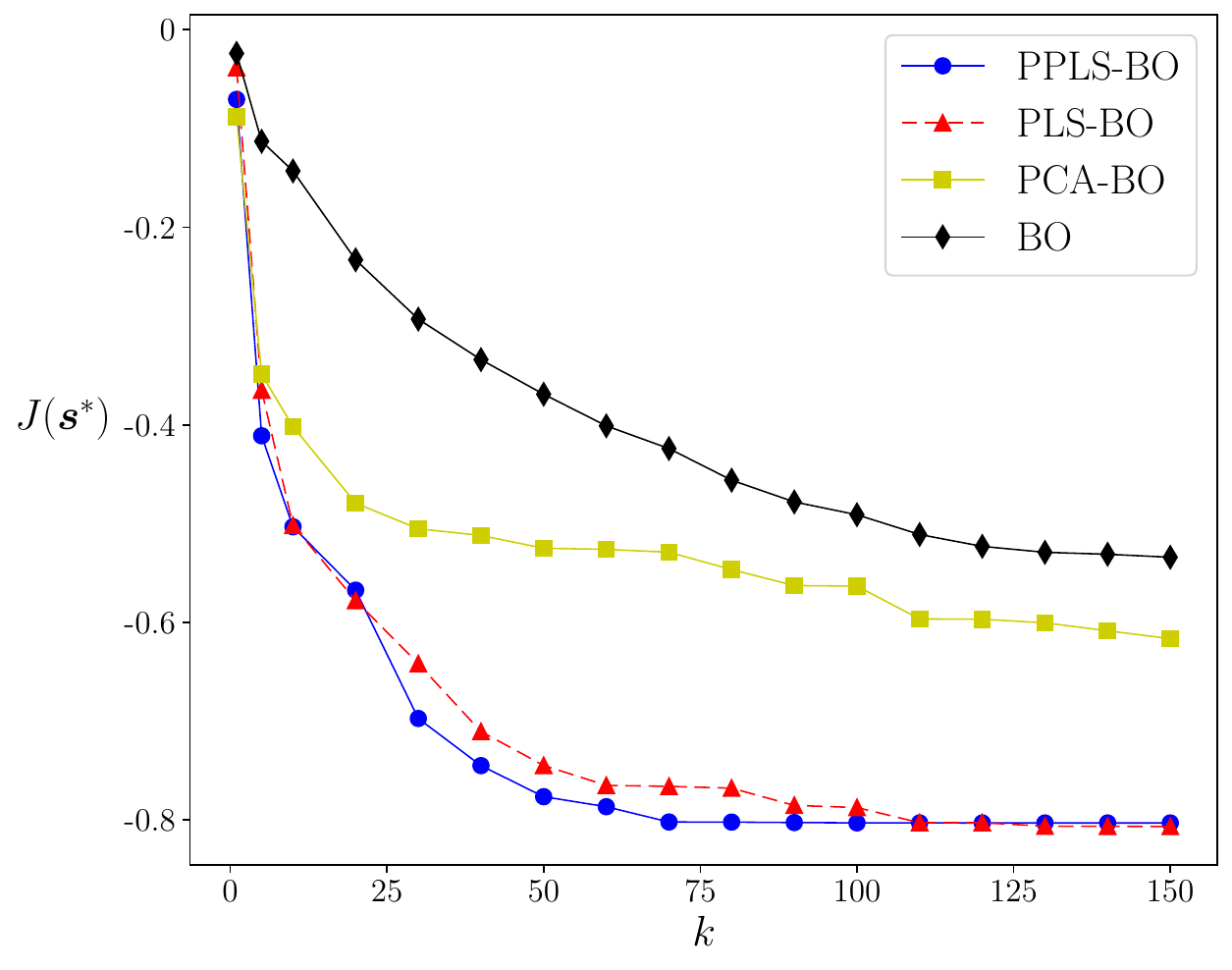}
	\caption{Illustrative example. Convergence of the optimum solution $J(\vec{s}^*)$ (mean) as a function of the number of adaptive sampling iterations $k$ comparing PPLS-BO, PLS-BO, PCA-BO and classical BO algorithms. PPLS-BO and PLS-BO use a latent variable dimension of $d_z=2$, while PCA-BO requires $d_z=8$.}
	\label{fig:ex1_conv}
\end{figure*}

When using the correct latent variable dimension $d_z = 2$, both PLS-BO and PPLS-BO algorithms propose solutions in close proximity of the global minimum; both converging to this solution within 100 adaptive sampling iterations. However, because PCA does not correctly discover the intrinsic dimension of the objective $J(\vec{s})$ and constraint $H(\vec{s})$ functions (Table~\ref{tab:ex1_init}), the latent space dimension must be increased to permit further exploration. A latent variable dimension of $d_z=8$ was found to be most effective for initialising PCA-BO to compare to PLS-BO and PPLS-BO. Slower convergence was observed for PCA-BO when compared to PLS-BO and PPLS-BO, although it improved on the solution computed using classical BO (Figure~\ref{fig:ex1_conv}).

This illustrative example demonstrates that PPLS-BO can be used to solve global minimisation problems that are impractical to solve when using classical BO. Furthermore, PPLS-BO adds additional flexibility such that the adaptive sampling is more robust to misspecification in the latent variable dimension $d_z$ when compared to the deterministic PLS-BO. Furthermore, the improved convergence of PLS-BO and PPLS-BO when compared to PCA-BO demonstrates that joint input-output training used in multi-view learning improves optimisation.

\subsection{Cantilever beam}

\begin{figure*}[t!]
	\centering
	\includegraphics[width=100mm]{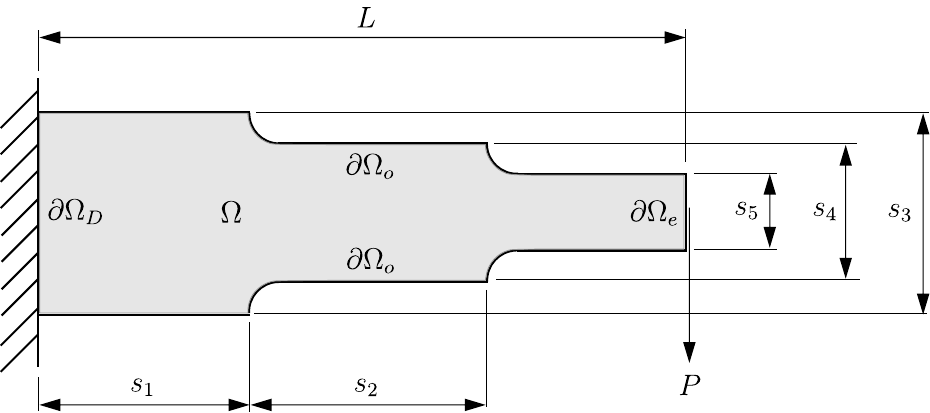}
	\caption{Cantilever beam schematic showing the domain $\Omega$ with Dirichlet boundary condition $\partial \Omega_D$ applied on the left edge, outer boundary $\partial \Omega_o$, the end boundary $\partial \Omega_e$ where the load $P$ is applied, and design variables $\vec{s}=(s_1\,s_2\,\cdots\,s_5)^\trans$ of the cantilever beam with total length $L$.}
	\label{fig:ex2_schematic}
\end{figure*}

Consider the cantilever beam depicted in  Figure \ref{fig:ex2_schematic} with a domain $\Omega$ discretised over spatial coordinates $\vec{x} \in \mathbb{R}^3$. The deformation measured by the displacement vector field, $\vec{g}(\vec{x},\vec{s}) \in \mathbb{R}^3$ is governed by the equations of (linear) elasticity, using the plane stress deformation assumption. The cantilever beam has an outer boundary $\partial \Omega_o$ and an end boundary $\partial \Omega_e$ where the load $P$ is applied. The Dirichlet boundary condition is applied along the left edge $\partial \Omega_D$, with the Neumann boundary condition applied along the remaining boundary $\partial \Omega_N$, where $\partial \Omega_N = \partial \Omega_o \cup \partial \Omega_e$.

\begin{figure*}[t!]
	\centering
	{\includegraphics[width=85mm]{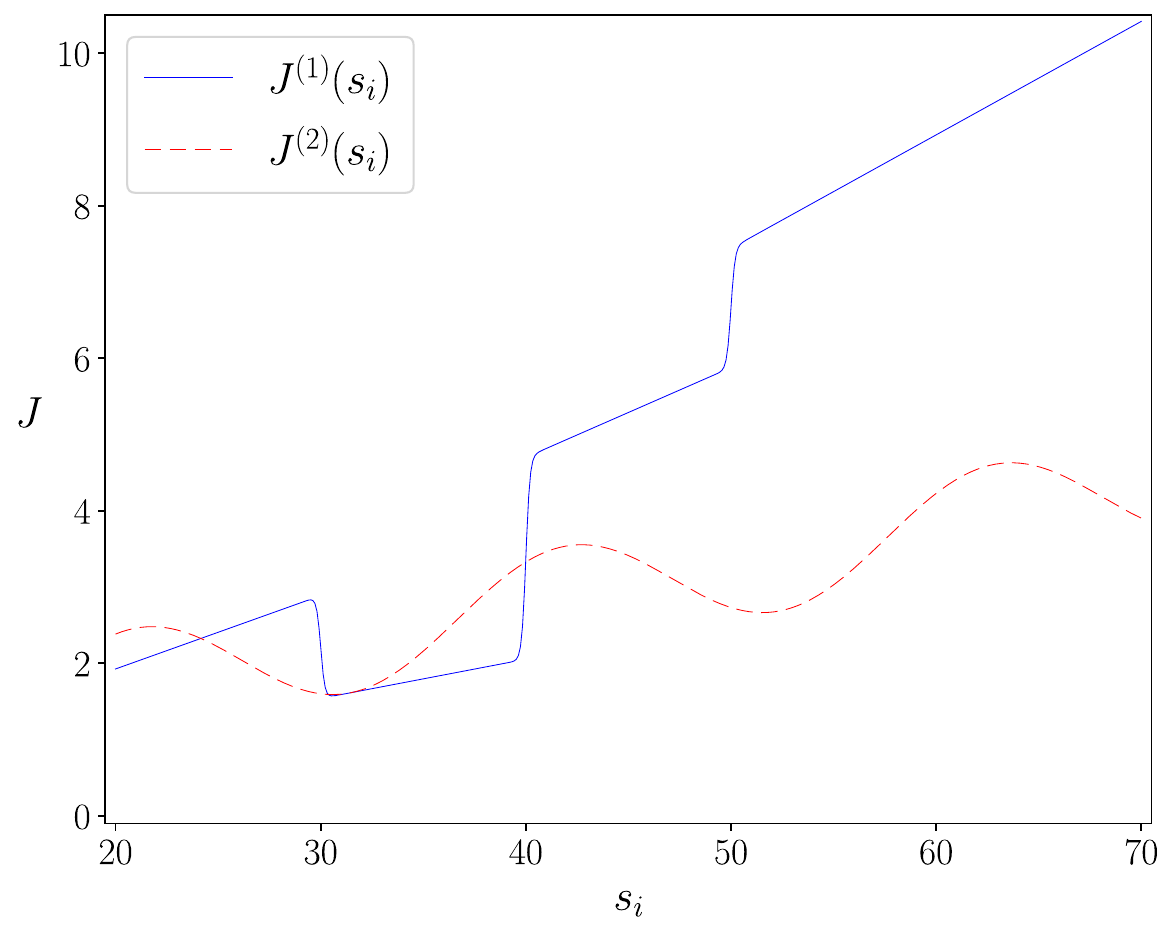}}
	\caption{\,Cantilever beam. Step $J^{(1)}(s_i)$ and periodic $J^{(2)}(s_i)$ objective functions.}
	\label{fig:ex2_j}
\end{figure*}

The geometry is parameterised by five design variables $\vec{s} = (s_1\,s_2\,\cdots\,s_5)^\trans$, where $s_1$ and $s_2$ control the length of each segment of the beam which has a total length $L=500$, and $s_3$, $s_4$, and $s_5$ control the depth of each segment. The body force equates to self-weight, which is constant over the geometry domain. The surface traction is zero along the outer edge $\partial \Omega_o$ and the resultant load $P$ is applied vertically along the end boundary $\partial \Omega_e$. The first two design variables \mbox{$s_i \in \{s_i \in \mathbb{R}\, \vert \, 100 \leq s_i \leq 200 \}$} for $i\in\{1,2\}$ and the remaining design variables \mbox{$s_i \in \{s_i \in \mathbb{R}\, \vert \, 20 \leq s_i \leq 70 \}$} for $i\in\{3,4,5\}$ are sampled from the domain $D_s$. In this example, we consider two "toy" objective functions $J^{(j)}(\vec{s})$, $j \in \{1,2\}$ each with different behavior. The goal is to compute the design variables that minimise the objective function
\begin{equation}\label{eq:ex2_3}
\begin{aligned}
J^{(j)}\left(\vec{s}\right) = J^{(0)}\left(\vec{s}\right) + \sum_{i=3}^5 J^{(j)}\left(s_i\right),
\end{aligned}
\end{equation}
where 
\begin{subequations} \label{eq:ex2_4}
\begin{align}
J^{(0)}\left(\vec{s}\right) &= 0.000108\left(s_1s_3 +s_2s_4 + s_5\left(500 - s_1 - s_2\right)\right),\label{eq:ex2_4a}\\
J^{(1)}\left(s_i\right) &= s_i\left(0.0963 - 0.0450 f_l\left(s_i - 30\right)+0.0662 f_l\left(s_i - 40\right) + 0.0313 f_l\left(s_i - 50\right)\right),\label{eq:ex2_4b}\\
J^{(2)}\left(s_i\right) &= 0.0513s_i + 1.38\cos^2\left(0.15s_i\right),\label{eq:ex2_4c}
\end{align}
\end{subequations}
and
\begin{equation}\label{eq:ex2_5}
\begin{aligned}
f_l(s_i) = \frac{1}{1 + e^{-100s_i}} \, ,
\end{aligned}
\end{equation}
subject to the displacement constraint
\begin{equation}\label{eq:ex2_6}
\begin{aligned}
H(\vec{s}) = \max_{x \in \Omega}\left(\vert\vert \vec{u}(\vec{x},\vec{s}) \vert\vert_2\right) - u_0,
\end{aligned}
\end{equation}
where $u_0=2$ is the displacement limit. The function $J^{(0)}(\vec{s})$ is approximately proportional to the volume of the domain $\Omega$ and $J^{(j)}(s_i)$ ($j \in \{1,2\}$) is an additional contribution inducing both step for $j=1$ and periodic for $j=2$ behaviours (Figure \ref{fig:ex2_j}). Although this is a "toy" example, these types of objective functions could represent the cost of different manufacturing processes where for example, the step function would represent certain supply chain constraints where it is relatively inexpensive to source components of specific sizes.

\begin{figure*}[t!]
\centering
\includegraphics[width=110mm]{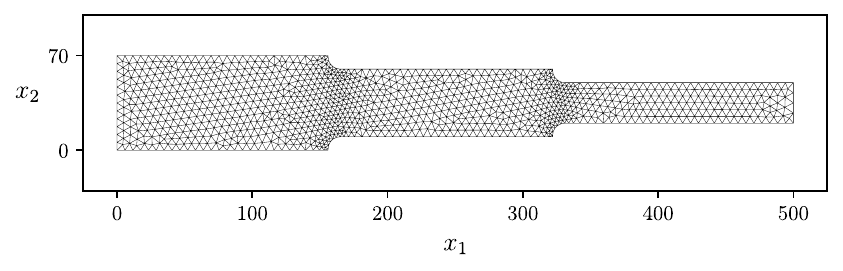}
\caption{\,Cantilever beam. FE mesh of an example geometry containing 2212 linear triangle elements, discretised over spatial coordinates $\vec{x} = (x_1\,x_2\,x_3)^\trans$ according to the plane stress assumption.}
\label{fig:ex2_mesh}
\end{figure*}

The constraint function $H(\vec{s})$ is sampled using the FE model shown in Figure \ref{fig:ex2_mesh}. The cantilever beam is modelled as a thin plate with a thickness of 5, using the plane stress assumption. A Young's modulus of $2\times10^{5}$  and Poisson's ratio of $0.3$ are used.

The adaptive sampling convergence to target solutions of 7.18 and 7.44 (as defined by the number of iterations required to propose a solution less than the target solution) for the step $J^{(1)}(\vec{s})$ and periodic $J^{(2)}(\vec{s})$ objective functions respectively for PPLS-BO is compared to PLS-BO and classical BO. The training data is initialised using $n=8$ orthogonal (PBD) samples. The two-level PBD uses design variables $\vec{s}$ with levels at their lower and upper limits. Convergence rates for both the UCB \eqref{eq:as_2} and EI \eqref{eq:as_3} constrained acquisition functions are compared. The convergence rates are computed across 10 runs for a mean and standard deviation. For the UCB acquisition function, an adaptive exploration parameter of $\gamma = 0.2d_z\ln(2(k+1))$ is used, and for the EI acquisition function, a jitter parameter of $\xi=0$ is used. An adaptive sampling budget of $n_k=100$ is used in all algorithms, with an EM budget of $n_t=100$ and MC sampling budget $n_l=10^3$ used in the PPLS-BO algorithm. A latent variable dimension of three ($d_z=3$) is selected for PPLS-BO and PLS-BO algorithms using cross-validation. This is compared to PCA-BO, which requires a latent variable dimension of four ($d_z = 4$).

For the step objective function $J^{(1)}(\vec{s})$, the convergence rates of PPLS-BO are on average between 2.3 and 3.2 times faster than classical BO. Convergence rates decrease for PLS-BO, which is between 1.5 and 1.7 times faster than classical BO. The convergence rates also exhibit less variance for PPLS-BO than for PLS-BO. PCA-BO only shows a modest improvement verses classical BO when compared to PPLS-BO and PLS-BO. The relative improvement is consistent across all data initialisation methods (where convergence is achieved) and acquisition functions. Although faster convergence is achieved on average when using PPLS-BO compared to PLS-BO, the variance in the convergence rates is high in all cases since the adaptive sampling is statistical in nature (Table \ref{tab:ex2_comp}). An optimum solution of $J^{(1)}(\vec{s}^*)=6.8$ is obtained with design variables \mbox{$\vec{s}^* = (129\,\, 200\,\, 32.0\,\, 32.1\,\, 32.8)^{\trans}$}.

For the periodic objective function $J^{(2)}(\vec{s})$, convergence rates of PPLS-BO are on average 5.7 to 6.2 times faster than classical BO. Convergence rates decrease for PLS-BO, which is between 4.4 and 5.3 times faster than classical BO. PCA-BO only shows a modest improvement verses classical BO when compared to PPLS-BO and PLS-BO. An optimum solution of $J^{(2)}(\vec{s}^*)=6.6$ is obtained with design variables \mbox{$\vec{s}^* = (133\,\,166\,\,31.6\,\,32.3\,\,29.6)^{\trans}$}.

Significantly more improvement in convergence rates between PPLS-BO, PLS-BO, and classical BO is observed for the periodic objective function $J^{(2)}(\vec{s})$ when compared to the step objective function $J^{(1)}(\vec{s})$. The gradient of the objective function within close proximity of the global minimum, is much smaller over a larger region of the design variable domain $D_s$ for the step objective function than for the periodic objective function. In this case, the global minimum can be more easily located using classical BO. Furthermore, GPs make smoothness assumptions based on the covariance function \eqref{eq:gp3}, which may poorly emulate the step objective function.

\begin{table*}[t!]
\begin{center}
\caption{\,Cantilever beam. Mean and standard deviation (in parenthesis) of the number of iterations $k$ required to satisfy convergence criteria averaged over 10 runs using PBD initialised data with UCB and EI constrained acquisition functions for the cost models $J^{(j)}(\vec{s})$.}
\begin{tabular*}{\textwidth}{@{\extracolsep{\fill}}cccccc}
\toprule
$J^{(j)}(\vec{s})$ & Acquisition Function & PPLS-BO & PLS-BO & PCA-BO  & BO \\\cmidrule{1-6}
\multirow{2}{*}{$J^{(1)}(\vec{s})$} & UCB & 11.4 (6.8) & 18.5 (8.3) & 26.2 (12.1) & 31.9 (14.2) \\
 &  EI & 13.4 (7.3) & 20.4 (11.9) & 26.3 (11.2) & 30.2 (13.0) \\\cmidrule{1-6}
 \multirow{2}{*}{$J^{(2)}(\vec{s})$} & UCB & 9.5 (4.7) & 13.4 (7.7) & 38.1 (19.3) & 58.7 (26.0) \\
  & EI & 10.1 (2.7) & 10.7 (6.5) & 37.9 (16.2) & 57.7 (21.5) \\\bottomrule
\end{tabular*}
\label{tab:ex2_comp}
\end{center}
\end{table*}

Since PPLS-BO proposes a probability density that design variables $\vec{s}$ are sampled from at each adaptive sampling iteration (as opposed to proposing a deterministic value with PLS-BO), this facilitates additional exploration, which leads to improvement in the rates of convergence. Furthermore, the basis $\vec{W}$ computed using PPLS-BO has a higher degree of sparsity in the linear combination of design variables $\vec{s}$, which further improves convergence by focusing exploration on the most influential design variables.

\begin{figure*}[b!]
	\centering
	\subfloat[Step objective function $J^{(1)}(\vec{s})$]{\includegraphics[width=70mm]{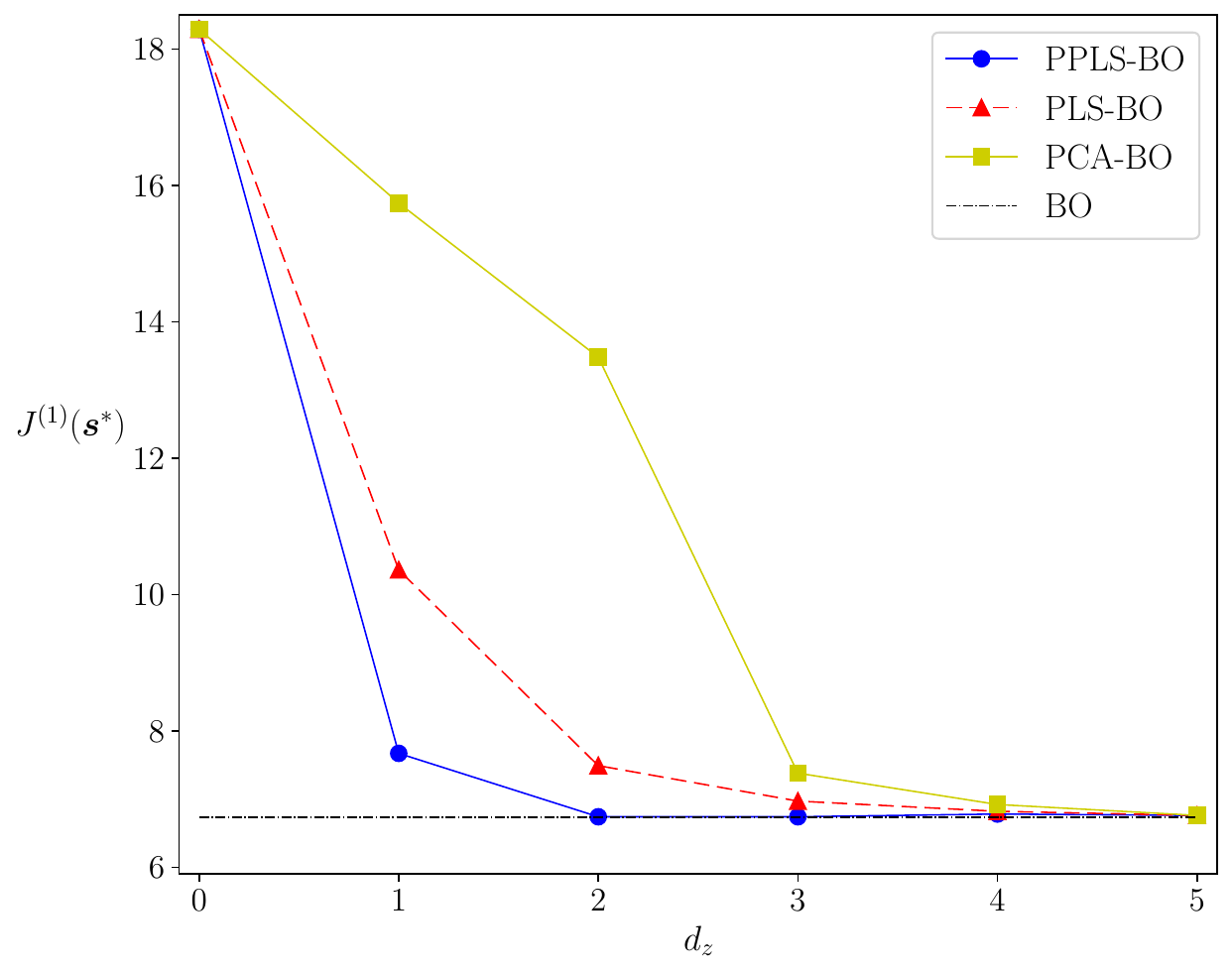}}
	\subfloat[Periodic objective function $J^{(2)}(\vec{s})$]{\includegraphics[width=70mm]{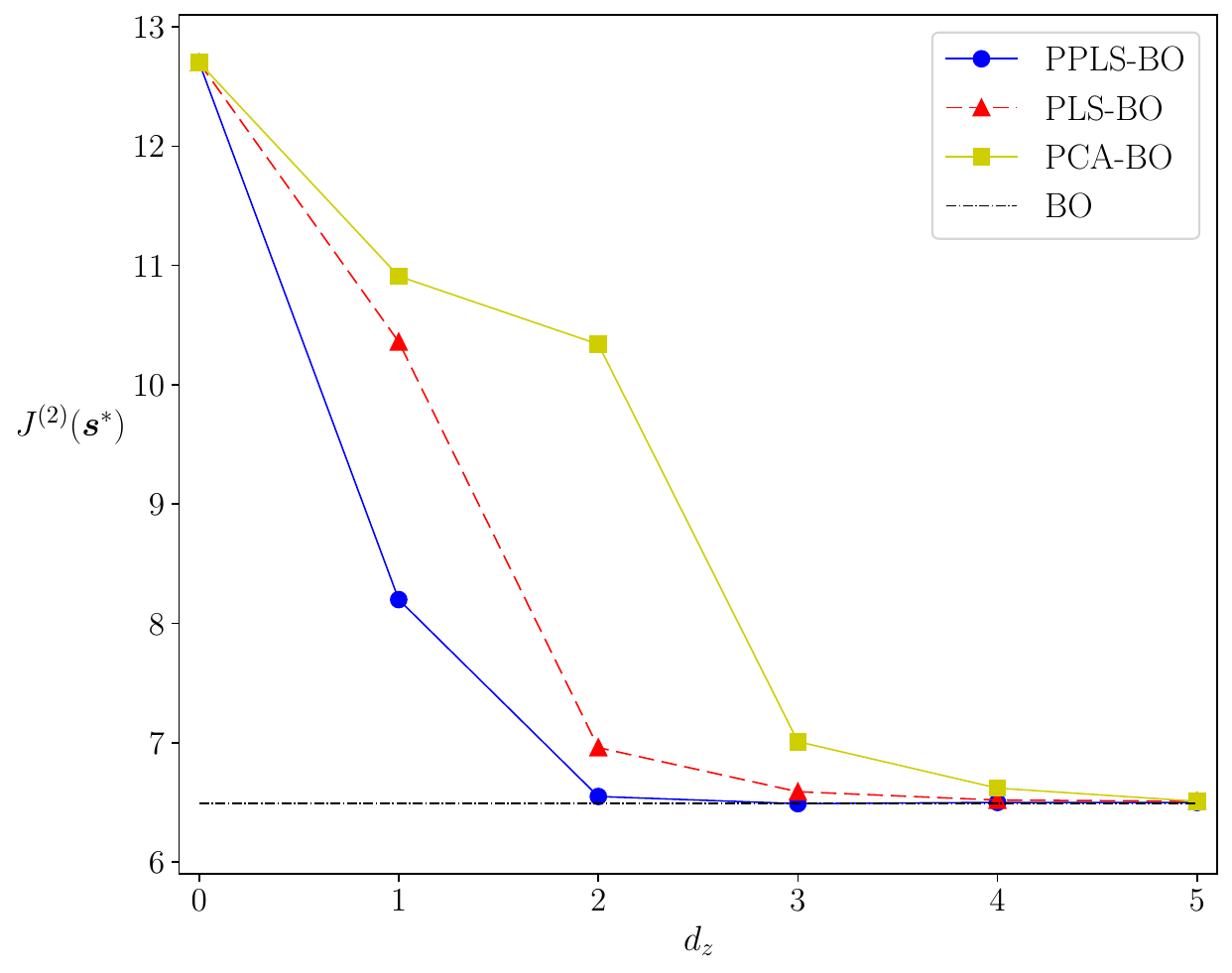}}
	\caption{\,Cantilever beam. Optimum solution $J^{(j)}(\vec{s})$ ($j\in\{1,2\}$) as a function of linear subspace dimension $d_z$ comparing PPLS-BO, PLS-BO, PCA-BO, and the classical BO benchmark solutions for the (a) step objective function $J^{(1)}(\vec{s})$, and (b) periodic objective function $J^{(2)}(\vec{s})$.}
	\label{fig:ex2_conv}
\end{figure*}

The properties of the PPLS-BO algorithm enhance exploration of the design variable domain $D_s$ when compared to PLS-BO, which is advantageous when the latent variable dimension $d_z$ is incorrectly specified. In the case of PLS-BO, the optimum solution for both the step $J^{(1)}(\vec{s})$ and periodic $J^{(2)}(\vec{s})$ objective functions is well approximated with a linear subspace dimension greater than three. However for PPLS-BO, the optimum solution is well approximated but with a linear subspace dimension of two. Consequently, PPLS-BO is more robust to misspecification of the linear subspace dimension $d_z$ when compared to PLS-BO (Figure \ref{fig:ex2_conv}). 

\subsection{Welded assembly}

\begin{figure*}[b!]
	\centering
	\subfloat[Boundary conditions]{\includegraphics[width=70mm]{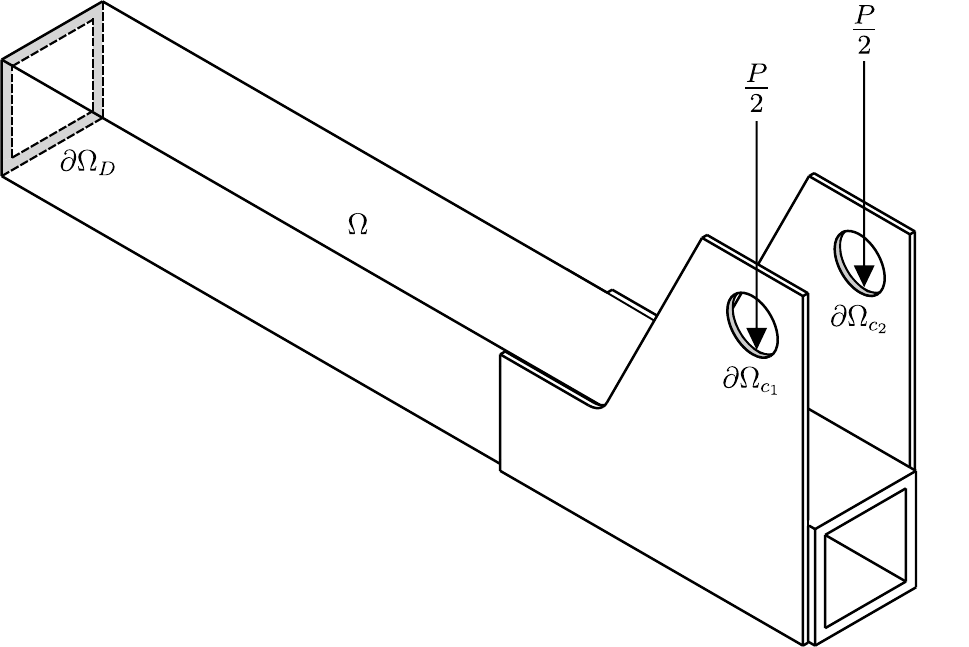}}
	\subfloat[Assembly components]{\includegraphics[width=70mm]{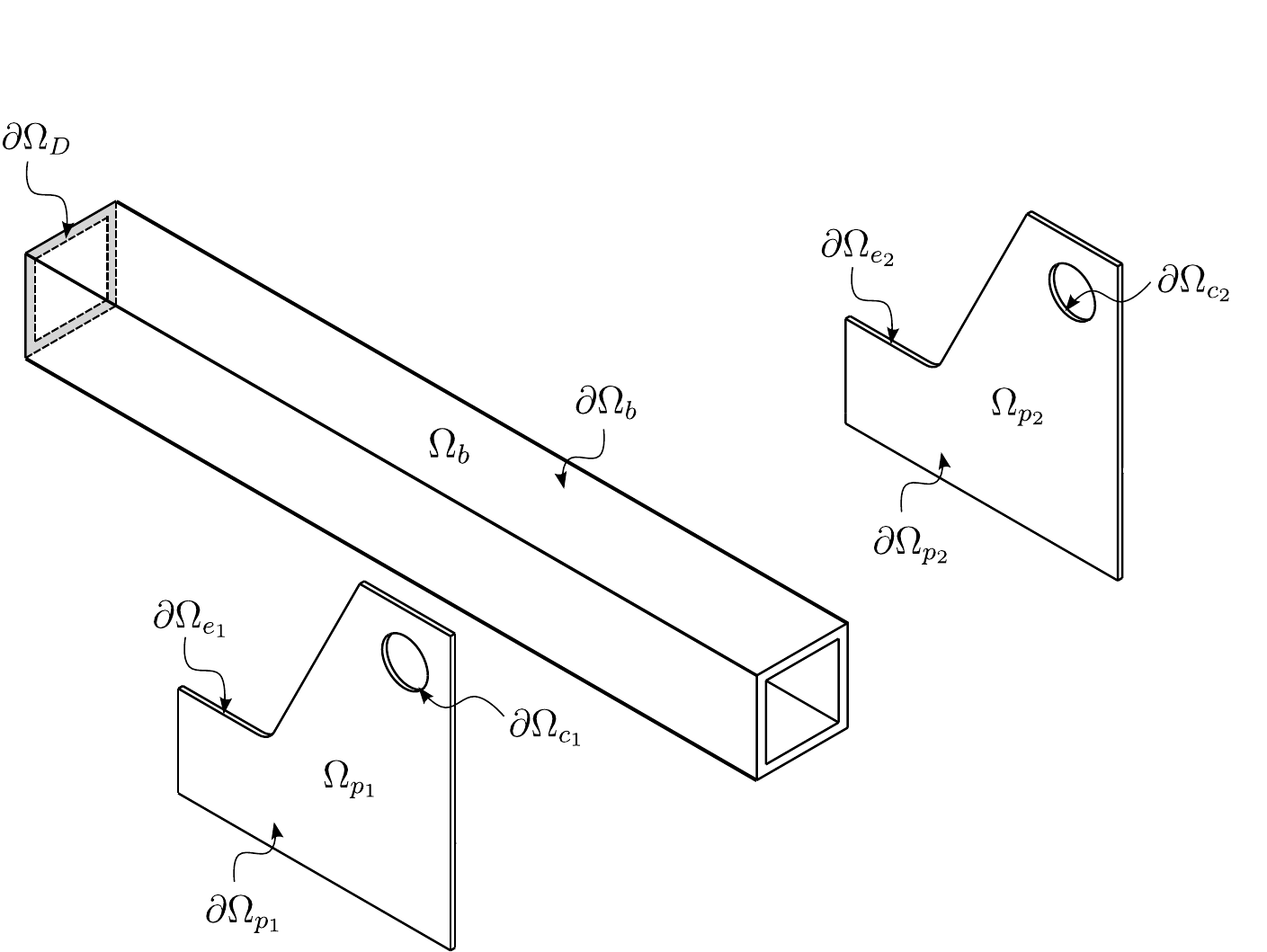}}
	\caption{Welded assembly. Schematic showing (a) the geometry domain $\Omega$ with left face boundary of the beam $\partial \Omega_D$ where the Dirichlet boundary condition is applied, and the circular hole face boundaries $\partial \Omega_{c_1}$ and $\partial \Omega_{c_2}$ where the load $P$ is applied. The (b) assembly components show the beam geometry $\Omega_b$ with boundaries $\partial \Omega_{b}$ and plates $\Omega_{p_1}$ and $\Omega_{p_2}$ with flat-area face boundaries $\partial \Omega_{p_1}$ and $\partial \Omega_{p_2}$, through-thickness edge face boundaries $\partial \Omega_{e_1}$ and $\partial \Omega_{e_1}$, and circular hole face boundaries $\partial \Omega_{c_1}$ and $\partial \Omega_{c_2}$.}
	\label{fig:ex3_geometry}
\end{figure*}

Consider the welded assembly shown in Figure~\ref{fig:ex3_geometry} with its deformation governed by the elasticity equations. The geometry represents a complex engineering component for a manufacturing application. The geometry domain $\Omega$ of the welded assembly is composed of a thin-walled beam $\Omega_b$ and two geometrically identical plates $\Omega_{p_1}$ and $\Omega_{p_2}$, such that $\Omega = \Omega_b \cup \Omega_{p_1} \cup \Omega_{p_2}$. All boundary faces of the thin-walled beam are denoted $\partial \Omega_b$ excluding the left boundary face $\partial \Omega_D$, which is where the Dirichlet boundary condition is applied. The geometry of the forward-most plate $\Omega_{p_1}$ for example, consists of through-thickness face boundaries $\partial \Omega_{e_1}$, flat-area face boundaries $\partial \Omega_{p_1}$, and the circular hole face boundaries $\partial \Omega_{c_1}$ where half of the total load $P = 10^4$ is applied. The second plate is defined similarly to the first. The boundary $\partial \Omega_N$ where the Neumann boundary condition is applied, is defined as the union of all faces of the geometry excluding the left face boundary of the beam $\partial \Omega_D$.

\begin{figure*}[b!]
	\centering
	{\includegraphics[width=140mm]{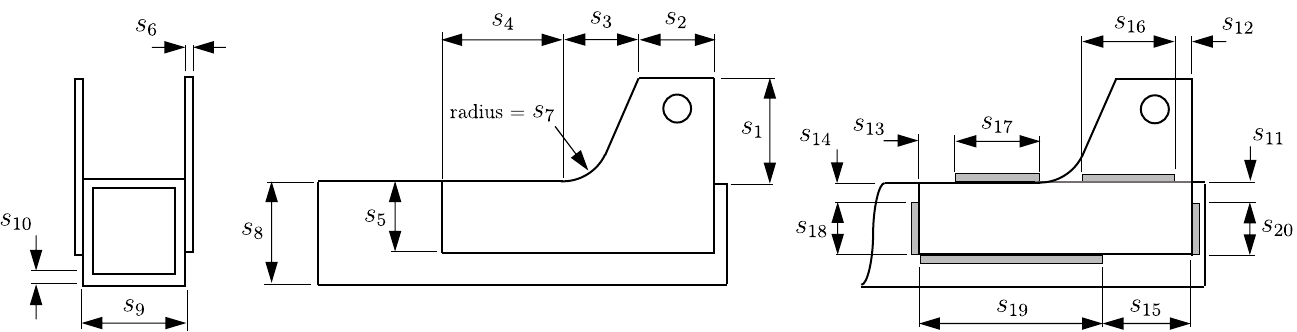}}
	\caption{Welded assembly. Schematic showing the design variables $s_i$, where $i \in \{1,2,\cdots,20\}$.}
	\label{fig:ex3_dimensions}
\end{figure*}

The domain of the welded assembly $\Omega$ is parameterised by 20 design variables $\vec{s} = (s_1\,\,s_2\, \cdots \, s_{20})^\trans$. Design variables $s_i$ for $i \in \{1,2,\cdots,7\}$ control the shape of the plates, design variables $s_i$ for $i \in \{8,9,10\}$ control the shape of the thin-walled beam, which has a fixed length $L=500$, design variables $s_i$ for $i \in \{11,12,\cdots,15\}$ control the position of the welds, and design variables $s_i$ for $i \in \{16,17,\cdots,20\}$ control the length of the welds, which have a total combined length $l_w$. All welds are positioned along the through-thickness faces $\partial \Omega_{e_1}$ and $\partial \Omega_{e_2}$ of each plate and intersect the beam faces $\partial \Omega_b$, excluding the weld of length $s_{16}$ which intersects the flat-area faces $\partial \Omega_{p_1}$ and $\partial \Omega_{p_2}$ and the beam face $\partial \Omega_b$ (Figure \ref{fig:ex3_dimensions}). The body force is constant over the geometry domain and is equal to self-weight. The surface traction is zero on all boundaries $\partial \Omega_N \backslash (\partial \Omega_{c_1} \cup \partial \Omega_{c_2})$ and yields half the resultant applied load along each of the circular hole boundaries. The design domain $D_s$ is defined by box-constraints for each design variable \mbox{$s_i \in \{s_i \in \mathbb{R}\, \vert \, \bar{s}_i^{(l)} \leq s_i \leq \bar{s}_i^{(u)} \}$} where $i \in \{1,2,\cdots,20\}$, and $\bar{s}_i^{(l)}$ and $\bar{s}_i^{(u)}$ are lower and upper limits on the design variables respectively (Table \ref{table:ex3_bounds}). In this example, the goal is to minimise manufacturing cost 

\begin{equation}\label{eq:ex3_1}
\begin{aligned}
J\left(\vec{s}\right) = J^{(0)}\left(\vec{s}\right) + J^{(1)}\left(\vec{s}\right) + 
J^{(2)}\left(\vec{s}\right),
\end{aligned}
\end{equation}
which may be formulated in terms of material $J^{(0)}\left(\vec{s}\right)$, welding $J^{(1)}\left(\vec{s}\right)$, and cutting $J^{(2)}\left(\vec{s}\right)$ costs~\cite{pavlovvcivc2004cost}. The material cost
\begin{equation}\label{eq:ex3_2}
\begin{aligned}
J^{(0)}\left(\vec{s}\right) = c_m\rho \scaleobj{.85}{\int_\Omega} d\vec{x},
\end{aligned}
\end{equation}
is computed based on a material cost per unit mass $c_m \in \mathbb{R}^+$, material density $\rho \in \mathbb{R}^+$, and the volume of material used. The welding operation cost
\begin{equation}\label{eq:ex3_3}
\begin{aligned}
J^{(1)}\left(\vec{s}\right) = c_w\left(t_w p_w(\vec{s}) l_w(\vec{s}) + t_{ws}\right),
\end{aligned}
\end{equation}
is computed using a welding cost per unit time $c_w \in \mathbb{R}^+$, welding time per unit length $t_w \in \mathbb{R}^+$, welding complexity coefficient $p_w(\vec{s})$ which accounts for additional time for more complex welds, total weld length $l_w(\vec{s})$, and welding setup time $t_{ws} \in \mathbb{R}^+$. For this example, the weld parameterised by length $s_{19}$ is more complex as it is an overhead weld so is assumed to take twice the welding time as the other welds, where
\begin{equation}\label{eq:ex3_4}
\begin{aligned}
 p_w(\vec{s}) =  1 + \frac{s_{19}}{l_w(\vec{s})}, \quad \quad l_w(\vec{s}) =  \sum_{i=16}^{20} s_i.
\end{aligned}
\end{equation}

\begin{table}[t!]
	\centering
	\caption{Welded assembly. Upper and lower limits for the box-constrained design variables \mbox{$s_i \in \{s_i \in \mathbb{R}\, \vert \, \bar{s}_i^{(l)} \leq s_i \leq \bar{s}_i^{(u)} \}$} where $i \in \{1,2,\cdots,20\}$.}\label{table:ex3_bounds}
	\scalebox{0.75}{
		\begin{tabular}{ccccccccccccccccccccc}
			\toprule
			& $s_1$ & $s_2$ & $s_3$ & $s_4$ & $s_5$ & $s_6$ & $s_7$ & $s_8$ & $s_9$ & $s_{10}$ & $s_{11}$ & $s_{12}$ & $s_{13}$ & $s_{14}$ & $s_{15}$ & $s_{16}$ & $s_{17}$ & $s_{18}$ & $s_{19}$ & $s_{20}$ \\\midrule
			$\bar{s}_i^{(l)}$ & 90 & 45 & 25 & 20 & 30 & 3 & 5 & 75 & 50 & 3 & 0 & 0 & 0 & 0 & 0 & 0 & 0 & 0 & 0 & 0\\
			$\bar{s}_i^{(u)}$ & 150 & 80  & 75 & 100 & 75 & 6 & 25 & 150 & 125 & 8 & 75 & 155 & 100 & 75 & 255 & 155 & 100 & 75 & 255 & 75\\
			\bottomrule
		\end{tabular}
	}
\end{table}

\noindent
The cutting operation cost
\begin{equation}\label{eq:ex3_5}
\begin{aligned}
J^{(2)}\left(\vec{s}\right) = c_c\left(p_c t_c(\vec{s}) l_c(\vec{s}) + t_{cs}\right),
\end{aligned}
\end{equation}
is computed using the cutting cost per unit time $c_c \in \mathbb{R}^+$, a cutting factor $p_c \in \mathbb{R}^+$ which scales for complexity (assumed constant in this example), cutting time per unit length of cut $t_c(\vec{s})$, total length of the cut $l(\vec{s})$, and cutting setup time $t_{cs} \in \mathbb{R}^+$. The cutting time per unit length $t_c(\vec{s})$ can be assumed to be proportional to the thickness~\cite{madic_co2_2018}, and the cutting length is based on the length of the outer edge of the identical plates, where
\begin{equation}\label{eq:ex3_6}
\begin{aligned}
t_c(\vec{s}) = \frac{1}{100}s_6, \quad \quad l_c(\vec{s}) = \frac{2}{s_6}\int_{\partial \Omega_{e_1} \cup \, \partial \Omega_{c_1}} \D\vec{x}.
\end{aligned}
\end{equation}
We assume a material cost per unit mass of $c_m = 1.5$, a welding cost per unit time of $c_w = 5.5 \times 10^{-3}$, a welding time per unit length of $t_w = 0.32$, a welding setup time of $t_{ws}=18$, a cutting cost per unit time of $c_c = 5 \times 10^{-3}$, a cutting factor of $p_c=1$, and a cutting setup time of $t_{cs}=180$. The manufacturing cost $J(\vec{s})$ is minimised subject to the displacement constraint
\begin{equation}\label{eq:ex3_7}
\begin{aligned}
H^{(1)}(\vec{s}) = \max_{x \in \Omega}\left(\vert\vert \vec{u}(\vec{x},\vec{s}) \vert\vert_2\right) - u_0,
\end{aligned}
\end{equation}
where $u_0=2$ is the limit on the total displacement, and the stress constraint
\begin{equation}\label{eq:ex3_8}
\begin{aligned}
H^{(2)}(\vec{s}) = \max_{x \in \Omega'_b}\left(\sigma_b(\vec{x},\vec{s})\right) - \sigma_{b0},
\end{aligned}
\end{equation}
where $\sigma_b(\vec{x},\vec{s})$ denotes the bending stress in the beam, and $\sigma_{b0}=125$ is the limit on the bending stress. An additional constraint is also imposed on the weld length
\begin{equation}\label{eq:ex3_9}
\begin{aligned}
H^{(3)}(\vec{s}) = l_{w0} - l_w(\vec{s}),
\end{aligned}
\end{equation}
such that the total length of weld $l_w(\vec{s})$ \eqref{eq:ex3_4} must exceed the minimum weld length $l_{w0}=100$.

The constraint functions $H^{(1)}(\vec{s})$ and $H^{(2)}(\vec{s})$ are sampled using an FE model, and are evaluated from any location in the beam geometry domain $\Omega_b$ excluding any location within four elements of the left boundary $\partial \Omega_D$ (this revised domain is denoted $\Omega'_b$). A Young's modulus of $2\times10^{5}$  and Poisson's ratio of $0.3$ are used. Since the geometry exhibits symmetry, only half of the geometry domain $\Omega$ is used to generate the FE model (reducing computational expense).

This is an example of a complex engineering component that is difficult to optimise with classical BO, due to the number of design variables $\vec{s}$. We compare the optimum manufacturing cost $J(\vec{s}^*)$ obtained using PPLS-BO and PLS-BO, within a fixed adaptive sampling budget $n_k$ for various latent variable dimensions $d_z$. We compare both algorithms with PCA-BO and classical BO. The data is initialised using PBD with $n=24$ samples, and adaptive sampling is performed using the EI acquisition function \eqref{eq:as_3} with constraints \eqref{eq:as_5} and a jitter parameter $\xi=0$. The two-level PDB uses the design variables $\vec{s}$ with levels at their lower and upper limits (Table \ref{table:ex3_bounds}). An adaptive sampling budget of $n_k=100$ is used with an EM budget of $n_t=100$ and MC sampling budget $n_l=10^3$ used in the PPLS-BO algorithm.

\begin{figure*}[b!]
\centering
{\includegraphics[width=80mm]{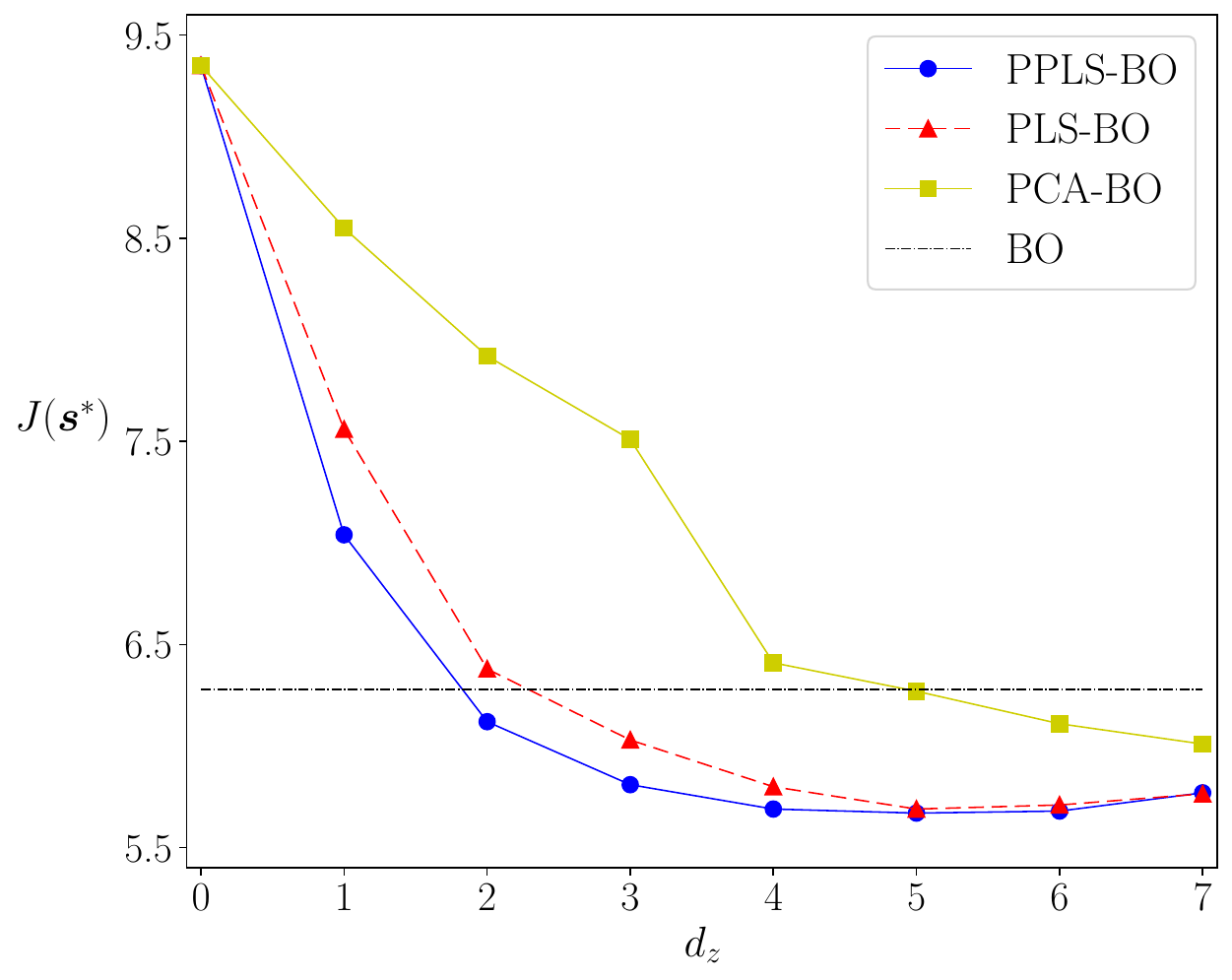}}
\caption{\,Welded assembly. Optimum manufacturing cost $J(\vec{s}^*)$ as a function of the latent variable dimension $d_z$ comparing PPLS-BO, PLS-BO, PCA-BO, and the classical BO benchmark solutions with a predetermined adaptive sampling budget of $n_k=100$.}
\label{fig:ex3_conv}
\end{figure*}

\begin{figure*}[t!]
\centering
\subfloat[von Mises stress $\sigma_v(\vec{x},\vec{s}^*)$]{\includegraphics[width=70mm]{ppls_bo_ex3_stress.pdf}}
\subfloat[Total displacement $\vert\vert \vec{u}(\vec{x},\vec{s}^*) \vert\vert_2$]{\includegraphics[width=70mm]{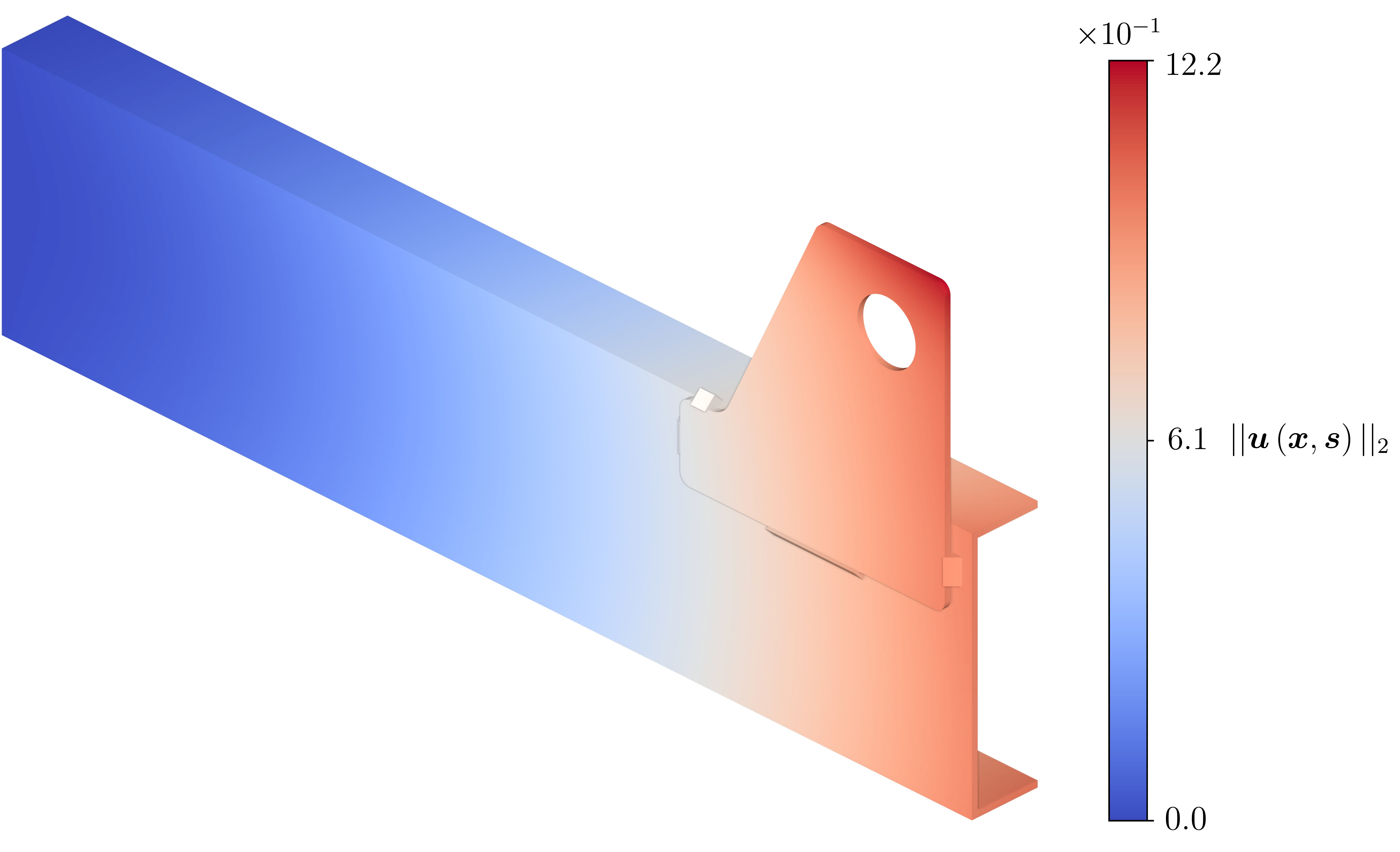}}
\caption{\,Welded beam. (a) von Mises stress $\sigma_v(\vec{x},\vec{s}^*)$ and (b) total displacement $\vert\vert \vec{u}(\vec{x},\vec{s}^*) \vert\vert_2$, as computed using the FE method using symmetry. The geometry is meshed using second-order tetrahedron elements, with a minimum of two elements through-thickness with the optimum design variables $\vec{s}^*$ computed using PPLS-BO with a linear subspace dimension of six $d_z=6$.}
\label{fig:ex3_results}
\end{figure*}

Both the PPLS-BO and PLS-BO algorithms obtain a more optimum manufacturing cost $J(\vec{s}^*)$ when compared to PCA-BO and classical BO (Figure \ref{fig:ex3_conv}), since classical BO scales poorly to problems with more than 10 design variables and PCA-BO does not explore the entirety of the space containing the problems intrinsic dimensions. The welded assembly objective $J(\vec{s})$ and constraint functions $H^{(1)}(\vec{s})$, $H^{(2)}(\vec{s})$, and $H^{(3)}(\vec{s})$, have a low intrinsic dimension since the material cost, displacement, and stress are predominantly influenced by the beam geometry. Consequently, a significant improvement from the mean geometry (computed when $d_z=0$) is obtained when using PPLS-BO and PLS-BO algorithms with three-dimensional latent variables where $d_z=3$. Additional improvement in the optimum manufacturing cost $J(\vec{s}^*)$ is obtained by further increasing the latent variable dimension $d_z$. The optimum welded assembly geometry exhibits a large flexural rigidity to satisfy the displacement and stress constraints while minimising manufacturing cost. For the optimum design variables $\vec{s}^*$, only the stress constraint is active. The weld lengths and positions minimally affect the manufacturing cost $J(\vec{s})$, displacement $H^{(1)}(\vec{s})$, and stress $H^{(2)}(\vec{s})$ constraint; consequently, the basis $\vec{W}$ promotes minimal exploration of the design variables relating to the welds, and their optimum values are approximately equal to the mean. The optimised component and its deflection and von Mises stress isocontours are shown in Figure \ref{fig:ex3_results}. 

An optimum latent variable dimension is obtained when the optimum manufacturing cost begins to diverge for any given adaptive sampling budget $n_k$. The optimum manufacturing cost begins to diverge with a latent variable dimension greater than six, i.e. $d_z>6$. As more dimensions are added for the same adaptive sampling budget $n_k$, the degree of exploration along the basis vectors (columns of the basis $\vec{W}$) composed of a linear combination of the most influential design variables $\vec{s}$ is reduced. Consequently, a trade-off exists between the latent variable dimension $d_z$ and the adaptive sampling budget $n_k$. As with the other examples considered in this paper, PPLS-BO can obtain similar optimum solutions but with a smaller latent variable dimension $d_z$ when compared to PLS-BO. 

The welded assembly example demonstrates how MVL is used to extend the application of BO to complex engineering components or systems that would otherwise be impractical to optimise with classical BO due to the large number of design variables. These complex engineering systems may have multiple constraints that must be considered when forming a low-dimensional basis, i.e.  $\vec{W}$, composed of linear combinations of the most influential design variables $\vec{s}$, describing an intrinsic dimensionality $d_z$ pertaining to the entire problem, not just the objective function $J(\vec{s})$. This ensures the linear subspace crosses the boundary between active and inactive regions of the constraint function, where the optimum solution exists.

\section{Conclusion}\label{sec:conclusion}

We introduced the novel PPLS-BO algorithm, which combines probabilistic partial least squares with Gaussian process surrogates to perform Bayesian optimisation in reduced dimension. Both the PPLS model and the  GP surrogate are trained sequentially. PPLS assumes a generative model for generating output data from unobserved low-dimensional random latent variables. The posterior probability density of the latent variables is computed by minimising the KL divergence using expectation maximisation. We marginalise the GP posterior predictive density over the random latent variables to obtain a mean and variance as inputs to a BO acquisition function. The acquisition function is then maximised to obtain a more optimal mean latent variable vector and subsequently sample the corresponding design variable from a conditional density, augmenting the training data. Overall, the PPLS-BO algorithm improves convergence rates compared to its deterministic counterparts, PLS-BO and classical BO. Furthermore, optimum solutions comparable to those computed using PLS-BO can be obtained with PPLS-BO but with fewer latent variable dimensions. This makes PPLS-BO more robust to incorrectly chosen latent variable dimensions and improves computational tractability in high-dimensional problems. 

However, when emulating functions that lack smoothness, the advantages of using PPLS-BO over PLS-BO and classical BO diminish. This is caused by shortcomings of the smoothness assumptions of GP covariance functions; therefore, exploring other surrogates such as random forests~\cite{breiman_random_2001} for problems involving discrete data or discontinuities is a promising direction for further research. Furthermore, other novel surrogates derived using variational Bayes~\cite{Archbold2024variational} could be further extended to adaptive sampling problems using a similar framework introduced in this paper. The expressivity of the GP surrogate may also be further enhanced by parameterising its mean and covariance with neural networks~\cite{rixner2021probabilistic,vadeboncoeur2023fully}. The independent learning of the low-dimensional subspace and the GP surrogate could enable further extensions, such as the learning of Riemannian manifolds, to account for more complex geometric constraints~\cite{jaquier2020high}.

\hypertarget{Appendix A}{\section*{Appendix A. Partial Least Squares}\label{sec:appendix_a}}

PLS is a family of MVL methods that attempt to obtain an orthonormal basis from one set of data that is highly correlated to an orthonormal basis obtained from the other data set~\cite{hoskuldsson1988pls}. PLS maximises the covariance between pairs of latent variables $\vec{z} \in \mathbb{R}^{d_z}$ and $\vec{v} \in \mathbb{R}^{d_v}$. The latent variables
\begin{subequations}\label{eq:pls_1}
\begin{align}
\vec{z} &= \vec{W}^\trans\vec{s},\label{eq:pls_1a}\tag{A.1a}\\
\vec{v} &= \vec{Q}^\trans\vec{y},\label{eq:pls_1b}\tag{A.1b} 
\end{align}
\end{subequations}
are low-dimensional linear combinations of the design variable vector $\vec{s}$ and output vector $\vec{y}$ respectively. The bases $\vec{W}$ and $\vec{Q}$ are elements of the Stiefel manifolds \mbox{$V_k(\mathbb{R}^{d_s}) = \{\vec{W} \in \mathbb{R}^{d_s \times d_z} \, \vert \,\vec{W}^\trans\vec{W} = \vec{I}\}$} and \mbox{$V_k(\mathbb{R}^{d_y}) = \{\vec{Q} \in \mathbb{R}^{d_y \times d_v} \, \vert \, \vec{Q}^\trans\vec{Q} = \vec{I}\}$} respectively. PLS maximised the covariance of design variable $\vec{S}$ and output data $\vec{Y}$ mean-centred matrices projected onto their respective bases, where
\begin{equation}\label{eq:pls_2}
\begin{aligned}
\max\limits_{W,Q} \quad &\trace\left(\vec{W}^\trans\vec{\Sigma}_{sy}\vec{Q}\right)\\
\textrm{s.t.} \quad &\vec{W}^\trans\vec{W} = \vec{Q}^\trans\vec{Q} = \vec{I},
\end{aligned}\tag{A.2}
\end{equation}
and $\vec{\Sigma}_{sy} = \vec{S}^\trans\vec{Y}$. This can be solved using the non-linear iterative partial least-squares (NIPALS) algorithm, which iteratively solves an eigenvalue problem for each basis vector~\cite{wold1975soft}. For each latent variable vector $\vec{z}$, the corresponding design variables
\begin{equation}\label{eq:pls_3}
\begin{aligned}
\vec{s} = \left(\vec{W}^\dagger\right)^\trans\vec{z} + \left(\vec{I} -\left(\vec{W}^\dagger\right)^\trans\vec{W}^\trans\right)\vec{e}
\end{aligned}\tag{A.3}
\end{equation}
can be computed, where $\vec{W}^\dagger \in \mathbb{R}^{d_z \times d_s}$ is the Moore-Penrose inverse of the basis $\vec{W}$, and $\vec{e} \in \mathbb{R}^{d_s}$ is the reconstruction error incurred by the low-rank approximation.

The PLS-BO algorithm for adaptive sampling is similar to classical BO (Algorithm \ref{alg:EGO}), except the PLS model is solved using NIPALS at each iteration, and the low-rank approximation is computed before solving the computational model to collect the corresponding observation vector. 

\hypertarget{Appendix B}{\section*{Appendix B. Derivation of Marginal Likelihood Covariance}\label{sec:appendix_b}}

The entries of the marginal likelihood covariance matrix $\hat{\vec{\Sigma}}_{ys}$ given by \eqref{eq:ppls_7}, can be derived using the properties of covariance. The input covariance is given by
\begin{equation}\label{eq:cov_1}
\scalebox{0.825}{$
\begin{aligned}
\cov(\vec{s},\vec{s}) &= \cov(\vec{W}\vec{z} + \hat{\vec{\epsilon}}_s,\vec{W}\vec{z} + \hat{\vec{\epsilon}}_s) = \cov(\hat{\vec{\epsilon}}_s,\hat{\vec{\epsilon}}_s)+ 2\vec{W}\cov(\vec{z},\hat{\vec{\epsilon}}_s) + \vec{W}\cov(\vec{z},\vec{z})\vec{W}^\trans = \hat{\vec{\Sigma}}_s + \vec{W}\vec{W}^\trans,
\end{aligned}\tag{B.1}
$}
\end{equation}
the output covariance is given by
\begin{equation}\label{eq:cov_2}
\scalebox{0.855}{$
	\begin{aligned}
	\cov(\vec{y},\vec{y}) &= \cov(\vec{Q}\vec{z} + \hat{\vec{\epsilon}}_y,\vec{Q}\vec{z} + \hat{\vec{\epsilon}}_y) = \cov(\hat{\vec{\epsilon}}_y,\hat{\vec{\epsilon}}_y)+ 2\vec{Q}\cov(\vec{z},\hat{\vec{\epsilon}}_y) + \vec{Q}\cov(\vec{z},\vec{z})\vec{Q}^\trans = \hat{\vec{\Sigma}}_y + \vec{Q}\vec{Q}^\trans,
	\end{aligned}\tag{B.2}
	$}
\end{equation}
and the cross covariance is given by
\begin{equation}\label{eq:cov_3}
\scalebox{0.775}{$
	\begin{aligned}
	\cov(\vec{s},\vec{y}) &=\cov(\vec{W}\vec{z} + \hat{\vec{\epsilon}}_s,\vec{Q}\vec{z} + \hat{\vec{\epsilon}}_y) = \vec{W}\cov(\vec{z},\vec{z})\vec{Q}^\trans + \vec{W}\cov(\vec{z},\hat{\vec{\epsilon}}_s) + \cov(\vec{z},\hat{\vec{\epsilon}}_y)\vec{Q}^\trans + \cov(\hat{\vec{\epsilon}}_s,\hat{\vec{\epsilon}}_y) = \vec{W}\vec{Q}^\trans.
	\end{aligned}\tag{B.3}
	$}
\end{equation}

\hypertarget{Appendix C}{\section*{Appendix C. Expectation Maximisation Algorithm}}

The EM algorithm is an iterative procedure based on alternating coordinate descent used to compute the optimum model hyperparameters $\vec{\Phi}^*$~\cite{moon1996expectation}. First we maximise with respect to the trial density $q(\vec{z})$ and then the PPLS hyperparameters $\vec{\Phi}$. Following from \eqref{eq:ppls_8} and \eqref{eq:ppls_14}, for fixed PPLS hyperparameters $\vec{\Phi}_t$ the optimum trial density is given by
\begin{equation}\label{eq:em_1}
\begin{aligned}
q_{t+1}(\vec{z}) &= \argmax_{q(\vec{z})} \int q(\vec{z}) \ln \left(\frac{p_{\Phi_t}(\vec{z}\vert\vec{y},\vec{s}) p_{\Phi_t}(\vec{y},\vec{s})}{q(\vec{z})}\right)\,\D\vec{z}\\
&= \argmax_{q(\vec{z})} \bigl( -D_{KL}\left(q(\vec{z}) \, \vert \vert \, p_{\Phi_t}(\vec{z}\vert\vec{y},\vec{s})\bigr) + \ln p_{\Phi_{t}}(\vec{y},\vec{s})\right)\\
&= p_{\Phi_t}(\vec{z}\vert\vec{y},\vec{s}).
\end{aligned}\tag{C.1}
\end{equation}
By fixing the trial density, the PPLS model hyperparameters can be computed by maximising
\begin{equation}\label{eq:em_2}
\begin{aligned}
\vec{\Phi}_{t+1} = \argmax_{\Phi} \expect_{ q_{t+1}(z)}\left( \ln p_{\Phi}(\vec{y},\vec{s}\vert \vec{z})\right).
\end{aligned}\tag{C.2}
\end{equation}
The first expectation term in the ELBO \eqref{eq:ppls_15} given by
\begin{equation} \label{eq:em_3}
\scalebox{0.825}{$
	\begin{aligned}
	\expect_{p_{\Phi_t}(z\vert y,s)} \left(\ln p_{\Phi_{t+1}}(\vec{y} , \vec{s} \vert \vec{z})\right) = - \frac{1}{2}\biggl( d_s\ln(2\pi) + \ln\vert (\hat{\vec{\Sigma}}_{s})_{t+1}\vert + \vec{s}^\trans (\hat{\vec{\Sigma}}_{s})_{t+1}^{-1}\vec{s} - 2\expect(\vec{z})^{\trans}\vec{W}_{t+1}^{\trans}(\hat{\vec{\Sigma}}_{s})_{t+1}^{-1}\vec{s} +\\  \trace\left(\expect\left(\vec{z}\vec{z}^\trans\right)\vec{W}_{t+1}^{\trans}(\hat{\vec{\Sigma}}_{s})_{t+1}^{-1}\vec{W}_{t+1}\right) + d_y\ln(2\pi) + \ln\vert (\hat{\vec{\Sigma}}_{y})_{t+1} \vert + \vec{y}^\trans(\hat{\vec{\Sigma}}_{y})_{t+1}^{-1}\vec{y} - \\2\expect(\vec{z})^{\trans}\vec{Q}_{t+1}^{\trans}(\hat{\vec{\Sigma}}_{y})_{t+1}^{-1}\vec{y} +  \trace\left(\expect\left(\vec{z}\vec{z}^\trans\right)\vec{Q}_{t+1}^{\trans}(\hat{\vec{\Sigma}}_{y})_{t+1}^{-1}\vec{Q}_{t+1}\right)\biggr).
	\end{aligned}\tag{C.3}
	$}
\end{equation}
is expanded using the likelihood $p_{\Phi_{t+1}}(\vec{y},\vec{s} \vert \vec{z}) = p_{\upsilon_{t+1}}(\vec{y}\vert\vec{z}) p_{\zeta_{t+1}}(\vec{s}\vert\vec{z})$. In the E-step of the EM algorithm, the expectation terms
\begin{equation} \label{eq:em_4}
\begin{aligned}
\expect(\vec{z}) = \left(\hat{\vec{\mu}}_z\right)_{t}, \quad \quad \expect\left(\vec{z}\vec{z}^\trans\right) = \left(\hat{\vec{\Sigma}}_z\right)_t + \expect(\vec{z}) \expect(\vec{z})^{\trans},
\end{aligned}\tag{C.4}
\end{equation}
are computed using the posterior probability density $p_{\Phi_{t}}(\vec{z} \vert \vec{y},\vec{s})$ for a known $\vec{\Phi}_t$ and laws of total variance. The expectation terms are pre-determined using PPLS model hyperparameters $\vec{\Phi}_{t}$ at the current time-step $t$.

In the M-step, the expectation term \eqref{eq:em_3} is maximised to determine the PPLS model hyperparameters for the subsequent iteration-step $\vec{\Phi}_{t+1}$, subject to bases $\vec{W}_{t+1}$ and $\vec{Q}_{t+1}$ being orthogonal. This can be written as an unconstrained optimisation problem with the Lagrangian
\begin{equation} \label{eq:em_5}
\scalebox{0.95}{$
	\begin{aligned}
	\mathcal{L}\left(\vec{\Phi}_{t+1}\right) = - \frac{1}{2}\biggl( d_s\ln(2\pi) + \ln\vert (\hat{\vec{\Sigma}}_{s})_{t+1}\vert + \vec{s}^\trans (\hat{\vec{\Sigma}}_{s})_{t+1}^{-1}\vec{s} - 2\expect(\vec{z})^{\trans}\vec{W}_{t+1}^{\trans}(\hat{\vec{\Sigma}}_{s})_{t+1}^{-1}\vec{s} +\\  \trace\left(\expect\left(\vec{z}\vec{z}^\trans\right)\vec{W}_{t+1}^{\trans}(\hat{\vec{\Sigma}}_{s})_{t+1}^{-1}\vec{W}_{t+1}\right) + d_y\ln(2\pi) + \ln\vert (\hat{\vec{\Sigma}}_{y})_{t+1} \vert + \vec{y}^\trans(\hat{\vec{\Sigma}}_{y})_{t+1}^{-1}\vec{y} -\\ 2\expect(\vec{z})^{\trans}\vec{Q}_{t+1}^{\trans}(\hat{\vec{\Sigma}}_{y})_{t+1}^{-1}\vec{y} +  \trace\left(\expect\left(\vec{z}\vec{z}^\trans\right)\vec{Q}_{t+1}^{\trans}(\hat{\vec{\Sigma}}_{y})_{t+1}^{-1}\vec{Q}_{t+1}\right)\biggr)-\\
	\trace\left(\left(\vec{W}_{t+1}^\trans\vec{W}_{t+1}-\vec{I}\right)\vec{\Lambda}_s\right) - \trace\left(\left(\vec{Q}_{t+1}^\trans\vec{Q}_{t+1}-\vec{I}\right)\vec{\Lambda}_y\right),
	\end{aligned}\tag{C.5}
	$}
\end{equation}
where $\vec{\Lambda}_s \in \mathbb{R}^{d_s \times d_s}$ and $\vec{\Lambda}_y \in \mathbb{R}^{d_y \times d_y}$ are diagonal matrices of Lagrange multiplier constants. Taking the derivative of the Lagrangian $\mathcal{L}(\vec{W}_{t+1},\vec{Q}_{t+1},(\hat{\vec{\Sigma}}_s)_{t+1},(\hat{\vec{\Sigma}}_y)_{t+1})$ with respect to $\vec{W}_{t+1}$ and $\vec{Q}_{t+1}$, and equating to zero yields
\begin{subequations} \label{eq:em_6}
\begin{align}
\vec{W}_{t+1} &= \vec{s}\expect(\vec{z})^{\trans}\left(\expect\left(\vec{z}\vec{z}^\trans\right) + \vec{\Lambda}_s\right)^{-1} \label{eq:em_6a}, \tag{C.6a}\\ 
\vec{Q}_{t+1} &= \vec{y}\expect(\vec{z})^{\trans}\left(\expect\left(\vec{z}\vec{z}^\trans\right) + \vec{\Lambda}_y\right)^{-1}\label{eq:em_6b} \tag{C.6b}.
\end{align}
\end{subequations}
However, since the bases are orthogonal, the following identity can be derived
\begin{subequations} \label{eq:em_7}
\begin{align}
\vec{W}_{t+1}^\trans\vec{W}_{t+1} = \vec{I} = \left(\left(\expect\left(\vec{z}\vec{z}^\trans\right) + \vec{\Lambda}_s\right)^{-1}\right)^\trans \expect(\vec{z}) \vec{s}^\trans \vec{s}\expect(\vec{z})^\trans\left(\expect\left(\vec{z}\vec{z}^\trans\right) + \vec{\Lambda}_s\right)^{-1} \label{eq:em_7a}, \tag{C.7a}\\ 
\vec{Q}_{t+1}^\trans\vec{Q}_{t+1} = \vec{I} = \left(\left(\expect\left(\vec{z}\vec{z}^\trans\right) + \vec{\Lambda}_y\right)^{-1}\right)^\trans \expect(\vec{z}) \vec{y}^\trans \vec{y}\expect(\vec{z})^\trans\left(\expect\left(\vec{z}\vec{z}^\trans\right) + \vec{\Lambda}_y\right)^{-1} \label{eq:em_7b}, \tag{C.7b}
\end{align}
\end{subequations}
which can be decomposed using the Cholesky decomposition
\begin{subequations}\label{eq:em_8}
\begin{alignat}{2}
\vec{L}_{W}\vec{L}_{W}^\trans &= \left(\expect\left(\vec{z}\vec{z}^\trans\right) + \vec{\Lambda}_s\right)^\trans\left(\expect\left(\vec{z}\vec{z}^\trans\right) + \vec{\Lambda}_s\right)\, &&= \expect(\vec{z})\vec{s}^\trans\vec{s}\expect(\vec{z})^\trans,
\label{eq:em_8a}\tag{C.8a}\\ 
\vec{L}_{Q}\vec{L}_{Q}^\trans  &= \left(\expect\left(\vec{z}\vec{z}^\trans\right) + \vec{\Lambda}_y\right)^\trans\left(\expect\left(\vec{z}\vec{z}^\trans\right) + \vec{\Lambda}_y\right)\, &&= \expect(\vec{z})\vec{y}^\trans\vec{y}\expect(\vec{z})^\trans\label{eq:em_8b}\tag{C.8b}.
\end{alignat}
\end{subequations}
The bases
\begin{subequations} \label{eq:em_9}
\begin{align}
\vec{W}_{t+1} &= \vec{s}\expect(\vec{z})^{\trans}\vec{L}_{W}^{-1} \label{eq:em_9a}, \tag{C.9a}\\ 
\vec{Q}_{t+1} &= \vec{y}\expect(\vec{z})^{\trans} 
\vec{L}_{Q}^{-1} \tag{C.9b}\label{eq:em_9b},
\end{align}
\end{subequations}
can therefore be computed using the Cholesky decomposition \eqref{eq:em_8a} and \eqref{eq:em_8b} respectively.

Similarly, the PPLS covariance matrix hyperparameters $(\hat{\vec{\Sigma}}_s)_{t+1}$ and $(\hat{\vec{\Sigma}}_y)_{t+1}$ can be computed by taking the derivative of the Lagrangian $\mathcal{L}(\vec{W}_{t+1},\vec{Q}_{t+1},(\hat{\vec{\Sigma}}_s)_{t+1},(\hat{\vec{\Sigma}}_y)_{t+1})$ with respect to the covariance matrix hyperparameters $(\hat{\vec{\Sigma}}_s)_{t+1}$ and $(\hat{\vec{\Sigma}}_y)_{t+1}$, and equating to zero, which yields
\begin{subequations} \label{eq:em_10}
\begin{align}
(\hat{\vec{\Sigma}}_s)_{t+1} &= \diag \left( \vec{s}\vec{s}^\trans - 2\vec{W}_{t+1}\expect(\vec{z})\vec{s}^\trans + \vec{W}_{t+1}\expect\left(\vec{z}\vec{z}^\trans\right)\vec{W}_{t+1}^\trans \right), \label{eq:em_10a} \tag{C.10a}\\
(\hat{\vec{\Sigma}}_y)_{t+1} &= \diag \left( \vec{y}\vec{y}^\trans - 2\vec{Q}_{t+1}\expect(\vec{z})\vec{y}^\trans + \vec{Q}_{t+1}\expect\left(\vec{z}\vec{z}^\trans\right)\vec{Q}_{t+1}^\trans \right). \label{eq:em_10b} \tag{C.10b}
\end{align}
\end{subequations}

The E and M steps can be repeated iteratively until the EM budget $n_t$ (which is chosen such that the PPLS model hyperparameters $\vec{\Phi}$ converge) is exceeded. The initial bases $\vec{W}_0$ and $\vec{Q}_0$ can be set using QR decomposition~\cite{golub2013matrix}, with the PPLS covariance matrix hyperparameters $(\hat{\vec{\Sigma}}_s)_{0}$ and $(\hat{\vec{\Sigma}}_y)_{0}$ set equal to the identity matrices (Algorithm \ref{alg:em}). Alternatively if implementing EM within the PPLS-BO algorithm, the initial bases and covariance matrix hyperparameters can be initialised using the model hyperparameters from the previous BO iteration.
\begin{algorithm}[h!]
\caption{EM}\label{alg:em}
\begin{algorithmic}
\setstretch{1.25}
\small
\State {\textbf{Data}: $\vec{S}\,,\vec{Y}$} \Comment{mean centred and normalised}
\State {\textbf{Input}: budget $n_{t}$ , linear subspace dimension $d_z$}
\State {Initialise $\vec{W}_0$ and $\vec{Q}_0$ with QR decomposition, $(\hat{\vec{\Sigma}}_s)_0 \leftarrow \vec{I}, (\hat{\vec{\Sigma}}_y)_0 \leftarrow \vec{I}$}
\For{$t \in \{0,1,\cdots,n_{t}-1\}$}
\State  {Compute expectations $\expect(\vec{z}), \, \expect\left(\vec{z}\vec{z}^\trans\right)$} \Comment{E-step \eqref{eq:em_4}}
\State {Compute bases \scalebox{0.9}{$\vec{W}_{t+1}$,$\vec{Q}_{t+1}$}, and covariance \scalebox{0.9}{$(\hat{\vec{\Sigma}}_s)_{t+1},(\hat{\vec{\Sigma}}_y)_{t+1}$}}\Comment{M-step \eqref{eq:em_9a},\eqref{eq:em_9b},\eqref{eq:em_10a},\eqref{eq:em_10b}}
\EndFor
\State {\textbf{Result}: $\vec{\Phi} \leftarrow \left\{\vec{W}_{t+1}, \vec{Q}_{t+1}, (\hat{\vec{\Sigma}}_s)_{t+1}, (\hat{\vec{\Sigma}}_y)_{t+1}\right\}$}
\end{algorithmic}
\end{algorithm}

\bibliographystyle{ama}
\bibliography{references}

\end{document}